\documentclass[12pt]{article}
\usepackage[dvips]{graphicx}

\begin{document}

\title{Renormalization of Hard-Core Guest Charges Immersed 
in Two-Dimensional Electrolyte}

\author{L. {\v S}amaj$^1$}

\maketitle

\begin{abstract}
This paper is a continuation of a previous one 
[L. {\v S}amaj, {\it J. Stat. Phys.} {\bf 120}:125 (2005)]
dealing with the renormalization of a guest charge immersed in 
a two-dimensional logarithmic Coulomb gas of pointlike $\pm$ unit charges, 
the latter system being in the stability-against-collapse regime 
of reduced inverse temperatures $0\le \beta <2$.
In the previous work, using a sine-Gordon representation of
the Coulomb gas, an exact renormalized-charge formula was derived 
for the special case of the {\em pointlike} guest charge $Q$, 
in its stability regime $\beta \vert Q\vert < 2$.
In the present paper, we extend the renormalized-charge treatment 
to the guest charge with a hard core of radius $\sigma$, 
which allows us to go beyond the stability border $\beta\vert Q\vert = 2$.
In the limit of the hard-core radius much smaller than the correlation length 
of the Coulomb-gas species and at a strictly finite temperature, 
due to the counterion condensation in the extended region 
$\beta\vert Q\vert >2$, the renormalized charge $Q_{\rm ren}$ 
turns out to be a periodic function of the bare charge $Q$ with period 1.
The renormalized charge therefore does not saturate at a specific finite 
value as $\vert Q\vert \to\infty$, but oscillates between two extreme values. 
In the high-temperature Poisson-Boltzmann scaling regime of limits 
$\beta\to 0$ and $Q\to\infty$ with the product $\beta Q$ being finite, 
one reproduces correctly the monotonic dependence of $\beta Q_{\rm ren}$ 
on $\beta Q$ in the guest-charge stability region $\beta\vert Q\vert <2$ and 
the Manning-Oosawa type of counterion condensation with the uniform 
saturation of $\beta Q_{\rm ren}$ at the value $4/\pi$ in the region 
$\beta\vert Q\vert\ge 2$.
\end{abstract}

\medskip

\noindent {\bf KEY WORDS:} Coulomb systems; logarithmic interactions;
sine-Gordon model; renormalized charge; counterion condensation. 

\vfill

\noindent $^1$ 
Institute of Physics, Slovak Academy of Sciences, D\'ubravsk\'a cesta 9, 
\newline 845 11 Bratislava, Slovak Republic; e-mail: fyzimaes@savba.sk

\newpage

\renewcommand{\theequation}{1.\arabic{equation}}
\setcounter{equation}{0}

\section{Introduction}
The concept of renormalized charge is of primary importance in the 
equilibrium statistical mechanics of colloids, see, e.g., refs.
\cite{Alexander,Lowen,Belloni,Levin,Trizac}.
This concept is based on the assumption that, at a finite temperature,
the electric potential induced by a ``guest'' (say colloidal) charged 
particle, immersed in an infinite electrolyte, exhibits, at large 
distances from this particle, basically the screening form given by
the high-temperature linear Debye-H\"uckel (DH) theory.
From the physical point of view, due to strong electrostatic interactions, 
the guest particle of bare electric charge $Q$ attracts oppositely charged 
electrolyte particles (counterions) to its immediate vicinity, and this 
decorated object may be considered as a new entity of lower 
renormalized charge $Q_{\rm ren}$.
The idea of renormalized charge was confirmed in the framework of
the nonlinear Poisson-Boltzmann (PB) approach \cite{Belloni,Trizac,Aubouy}: 
there is a renormalized-charge prefactor, $Q_{\rm ren}$, to the usual Yukawa 
decay which is different from the bare charge $Q$ of the guest particle.   
An interesting point is that, as the absolute value of $Q$ increases to
infinity, the renormalized charge saturates monotonically at some finite 
value $Q_{\rm ren}^{\rm sat}$.
The possibility of a more general phenomenon of {\em potential} saturation
was studied from a general point of view in ref. \cite{Tellez03} and on
the exactly solvable 2D Coulomb gas at the Thirring free-fermion
point in ref. \cite{Samaj05a}.

The rigorous validity of the PB approach is restricted, under certain 
conditions making the nonlinear PB theory superior to the linear DH one
\cite{Kennedy1}, to a specific scaling regime of the infinite-temperature 
limit \cite{Kennedy2}.
To go beyond this mean-field description, one has to incorporate
electrostatic correlations among the electrolyte particles,
like it was done, e.g., in refs. \cite{Groot,Barbosa}.
Such approaches always involve some plausible, but not rigorously
justified, arguments and approximation schemes.

Another strategy is to concentrate on simplified models which keep 
the Coulomb nature of particle interactions and simultaneously admit 
the exact solution.
Specific two-dimensional (2D) Coulomb systems with logarithmic pairwise 
interactions among charged constituents, where the electrolyte is modelled 
by an infinite symmetric Coulomb gas, belong to such category of models.
The 2D Coulomb gas of $\pm$ unit pointlike charges is stable against
the thermodynamic collapse of positive-negative pairs of charges
at high enough temperatures, namely for $\beta<2$ with $\beta$ being
the (dimensionless) inverse temperature.  
In this stability region, the equilibrium statistical mechanics of 
the Coulomb gas is exactly solvable via an equivalence with
the integrable (1+1)-dimensional sine-Gordon theory; for a short review,
see ref. \cite{Samaj03}.
An extension of the exact treatment of the stable 2D Coulomb gas
to the presence of some {\em pointlike} guest charge(s) was done
in ref. \cite{Samaj05b}.
The problem of one (two) guest charge(s) immersed in the 2D Coulomb plasma
was shown to be related to the evaluation of one-point (two-point)
expectation values of the exponential field in the equivalent
sine-Gordon model.
Based on recent progress in the latter topic, two main results
were obtained.
Firstly, an explicit formula for the chemical potential of single
guest charge $Q$ was found in the guest-charge stability (no collapse
of $Q$ with a unit plasma counterion) region $\beta \vert Q\vert <2$.
Secondly, the asymptotic large-distance behavior of an effective
interaction between two guest charges $Q$ and $Q'$ was derived.
As a by-product of this result, considering that $Q'$ corresponds
to either $+1$ or $-1$ charged Coulomb-gas species, the concept 
of renormalized charge was confirmed and the explicit dependence of 
$Q_{\rm ren}$ on $\beta$ and $Q$, valid rigorously in the whole 
stability range of $\vert Q\vert < 2/\beta$, was established.
For a fixed $\beta<2$, the renormalized charge $Q_{\rm ren}$,
considered as a function of, say positive, bare charge $Q$,
exhibits a maximum at $Q=2/\beta-1/2$: this non-monotonic behavior
resembles the one observed in the Monte-Carlo simulations of
the salt-free (only counterions are present) colloidal cell
model \cite{Groot}.
At the stability border $Q=2/\beta$, $Q_{\rm ren}$ attains a finite value.
This fact was an inspiration for a conjecture, referred to as
``regularization hypothesis'', about the possibility of an analytic
continuation of the formula for $Q_{\rm ren}$ to the collapse
region $Q\ge 2/\beta$ \cite{Samaj05b}.

The validity of the regularization hypothesis was put in doubts
by T\'ellez \cite{Tellez05a} who calculated in detail the formula
for the renormalized charge within the 2D nonlinear PB theory.
The PB approach describes correctly, under certain conditions \cite{Kennedy2}, 
the scaling regime of limits $\beta\to 0$ and $Q\to\infty$ with
the product $\beta Q$ being finite.
As was expected, in the guest-charge stability region $\beta\vert Q\vert <2$, 
the exact formula for $Q_{\rm ren}$ \cite{Samaj05b}, taken with $\beta\to 0$,
was reproduced.
When $\beta\vert Q\vert \ge 2$, a hard core of radius $\sigma>0$, impenetrable
for electrolyte particles, has to be attached to the guest charge $Q$ 
in order to prevent its collapse with electrolyte counterions.
For the determination of the renormalized charge in this regularized case,
the connexion problem of the PB equation to relate the large-distance
behavior of the induced electric potential with its short-distance
expansion \cite{Tracy} is of primary importance.  
The numerical PB results of ref. \cite{Tellez05a} show that
$Q_{\rm ren}$ is always an increasing function of the bare $Q$,
and saturates at a finite value as $Q\to\infty$.
In particular, when the dimensionless positive hard-core radius
${\hat\sigma} = \kappa \sigma$ ($\kappa$ denotes the inverse Debye length)
is very small, ${\hat\sigma}\to 0$, 
the renormalized charge saturates at the value given by
\begin{equation} \label{1.1}
\beta Q_{\rm ren}^{\rm (sat)} = \frac{4}{\pi} \quad
\mbox{for {\em all} values of $\beta Q \ge 2$.}
\end{equation}
This is a manifestation of the Manning-Oosawa counterion condensation 
\cite{Manning,Oosawa} known in the theory of cylindrical polyelectrolytes.

The study of the renormalization charge within the nonlinear PB approach
\cite{Tellez05a} is trustworthy, however, there are two open problems.
Firstly, the rigorous validity of the PB theory was proved for 3D
electrolytes in the presence of some {\em continuous} domains of 
{\em external} charge distributions \cite{Kennedy2}.
The hard-core interaction between the external guest charge and the
Coulomb-gas particles, which is so relevant in the questionable
guest-charge collapse region $\beta\vert Q\vert \ge 2$, was not considered
in Kennedy's proof.
The second problem, which is probably even more important, is related to
the limit $\beta\to 0$ considered in the PB approach.
This limit represents a very strong restriction which prevents one
from seeing nontrivialities in the plot of $Q_{\rm ren}$ versus $Q$
appearing at a strictly finite (nonzero) $\beta$, like the previously mentioned
existence of the maximum extreme at the point $Q=2/\beta-1/2$ \cite{Samaj05b}.
It is evident that the monotonic increase of $\beta Q_{\rm ren}$ as 
the function of $\beta Q$ to its saturation value, predicted by the PB theory, 
is certainly not true for a finite inverse temperature $\beta$. 

The aim of the present paper is to extend the exact renormalized-charge
treatment of ref. \cite{Samaj05b} to the guest charge with a hard core
of radius $\sigma$, which allows us to go beyond the stability border
$\beta \vert Q\vert =2$ of the pointlike guest charge.
By the spirit, the applied method is similar to the one used in 
ref. \cite{Kalinay} to include hard cores around charged species of 
the infinite 2D Coulomb gas itself. 
In the limit ${\hat\sigma}\to 0$, i.e., when the hard-core radius of the guest 
particle is much smaller than the mean interparticle distance of 
the Coulomb-gas species, due to the counterion condensation in the extended 
region $\beta\vert Q\vert >2$, the renormalized charge $Q_{\rm ren}$ 
turns out to be a periodic function of the bare charge $Q$ with period 1. 
The renormalized charge therefore does not saturate at a specific finite 
value as $\vert Q\vert \to\infty$, but oscillates between two extreme values. 
Such behavior indicates that, for a strictly nonzero $\beta$,
the Manning-Oosawa counterion condensation phenomenon \cite{Manning,Oosawa}
should be revisited.
In the scaling PB regime, one recovers correctly the standard results 
including the saturation formula (\ref{1.1}).

The paper is organized as follows.
Section 2 reviews the known exact information about the equilibrium 
statistical mechanics of the infinite 2D Coulomb gas, including 
the complete thermodynamics and both short- and large-distance asymptotic 
behaviors of two-point correlation functions.
In this section, we introduce the notation and present important formulas 
which are extensively used throughout the whole paper.
Section 3 deals with the chemical potential of the guest charge immersed
in the 2D Coulomb gas.
The case of the pointlike guest charge is analyzed in Section 3.1, 
the inclusion of the hard-core region around the guest charge is
the subject of Section 3.2.
The renormalization of the guest charge is studied in Section 4.
In Section 4.1, the renormalization of the pointlike guest charge
is briefly reviewed following the derivation of ref. \cite{Samaj05b}.
In Section 4.2, relevant ${\hat\sigma}$-corrections due to the
presence of the hard core are systematically generated in the formula for the
renormalized charge.
A recapitulation is given in Section 5.

\renewcommand{\theequation}{2.\arabic{equation}}
\setcounter{equation}{0}

\section{Bulk properties of the 2D Coulomb gas}

\subsection{Integrability}
We consider a classical (i.e. non quantum) Coulomb gas consisting of 
two species of pointlike particles with opposite unit charges
$q\in \{ +1, -1 \}$, constrained to an infinite 2D plane $\Lambda$ 
of points ${\bf r} \in R^2$.
The interaction energy of a set of particles $\{ q_j, {\bf r}_j \}$ 
is given by $\sum_{j<k} q_j q_k v(\vert {\bf r}_j-{\bf r}_k\vert)$, 
where the Coulomb potential $v({\bf r}) = - \ln(\vert {\bf r}\vert/r_0)$
(the free length constant $r_0$ will be set to unity for simplicity)
is the regular solution of the 2D Poisson equation
$\Delta v({\bf r}) = - 2 \pi \delta({\bf r})$.
The equilibrium statistical mechanics of the system is usually
treated in the grand canonical ensemble, characterized by
the dimensionless inverse temperature $\beta$ and by the couple
of particle fugacities $z_+$ and $z_-$.
Since the length scale $r_0$ was set to unity, the true dimension
of $z_{\pm}$ is $[{\rm length}]^{-2+(\beta/2)}$.
The bulk Coulomb gas is neutral, and thus its thermodynamic properties 
depend only on the combination $\sqrt{z_+ z_-}$ \cite{Lieb}.
It is therefore possible to set $z_+=z_-=z$; however, at some places, 
in order to distinguish between the $+$ and $-$ charges, we shall
keep the notation $z_{\pm}$.
The system of pointlike particles is stable against the UV collapse 
of positive-negative pairs of unit charges provided that the corresponding 
Boltzmann weight $\exp[\beta v({\bf r})] = \vert {\bf r} \vert^{-\beta}$ 
can be integrated at short distances in 2D, i.e.,  at high enough
temperatures such that $\beta<2$.
In what follows, we shall restrict ourselves to this stability range 
of inverse temperatures. 

The grand partition function $\Xi$ of the 2D Coulomb gas can be
turned via the Hubbard-Stratonovich transformation 
(see, e.g., ref. \cite{Minnhagen}) into 
\begin{equation} \label{2.1}
\Xi(z) = \frac{\int {\cal D}\phi \exp[-S(z)]}{\int {\cal D}\phi \exp[-S(0)]}
\end{equation}
with
\begin{equation} \label{2.2}
S(z) = \int_{\Lambda} {\rm d}^2 r \left[ \frac{1}{16\pi} (\nabla\phi)^2
- 2 z \cos( b\phi ) \right] , \quad b = \sqrt{\frac{\beta}{4}}
\end{equation}
being the Euclidean action of the $(1+1)$-dimensional sine-Gordon model. 
Here, $\phi({\bf r})$ is a real scalar field and $\int {\cal D}\phi$
denotes the functional integration over this field. 
The sine-Gordon coupling constant $b$ depends only on
the inverse temperature $\beta$ of the Coulomb gas.
The fugacity $z$ is renormalized by the (diverging) self-energy term
$\exp[\beta v({\bf 0})/2]$ which disappears from statistical relations
under the conformal short-distance normalization of the exponential fields
\begin{equation} \label{2.3}
\langle {\rm e}^{{\rm i}b\phi({\bf r})} {\rm e}^{-{\rm i}b\phi({\bf r}')}
\rangle \sim \vert {\bf r}-{\bf r}'\vert^{-4 b^2}
\quad \mbox{as $\vert {\bf r}-{\bf r}' \vert \to 0$.}
\end{equation}

For $b^2<1$ ($\beta<4$), the discrete symmetry
$\phi\to \phi + 2\pi n/b$ ($n$ being an integer) of the
sine-Gordon action (\ref{2.2}) is spontaneously broken and therefore
the sine-Gordon model is massive \cite{Zamolodchikov79}. 
Its particle spectrum consists of one soliton-antisoliton pair
$(S,{\bar S})$ with equal masses $M$, which coexist in pairs, and of 
$S-{\bar S}$ bound states, called ``breathers'' 
$\{ B_j; j=1,2,\cdots < 1/\xi \}$, whose quantized number 
at a given $b^2$ depends on the inverse of the parameter
\begin{equation} \label{2.4}
\xi = \frac{b^2}{1-b^2} \quad \left( = \frac{\beta}{4-\beta} \right) . 
\end{equation}
The mass of the $B_j$ breather is given by
\begin{equation} \label{2.5}
m_j = 2 M \sin \left( \frac{\pi \xi}{2} j \right) 
\end{equation}
and this breather disappears from the sine-Gordon particle
spectrum just when $m_j = 2 M$, i.e., $\xi = 1/j$.
Note that the lightest $B_1$ breather is present in the spectrum
up to the collapse point $b^2=1/2$ $(\beta=2)$.

The 2D sine-Gordon model is an integrable field theory, so that any
multiparticle scattering $S$-matrix factorizes into a product
of explicitly available two-particle $S$-matrices satisfying
the Yang-Baxter identity \cite{Zamolodchikov79}. 
Basic characteristics of the underlying theory were derived quite 
recently by using the method of Thermodynamic Bethe ansatz.
In particular, the (dimensionless) specific grand potential $\omega$,
defined by
\begin{equation} \label{2.6}
- \omega = \lim_{\vert \Lambda\vert \to \infty} 
\frac{1}{\vert\Lambda\vert} \ln \Xi(z) ,
\end{equation}
was found in ref. \cite{Destri}:
\begin{equation} \label{2.7}
- \omega = \frac{m_1^2}{8 \sin(\pi\xi)} .
\end{equation}
Under the conformal normalization of the exponential fields (\ref{2.3}), 
the relationship between the fugacity $z$ and the soliton/antisoliton 
mass $M$ reads \cite{Zamolodchikov95}
\begin{equation} \label{2.8}
z = \frac{\Gamma(b^2)}{\pi \Gamma(1-b^2)}
\left[ M \frac{\sqrt{\pi} \Gamma((1+\xi)/2)}{2 \Gamma(\xi/2)}
\right]^{2(1-b^2)} ,
\end{equation}
where $\Gamma$ stands for the Gamma function.

\subsection{One-point densities}
The homogeneous number density of the Coulomb-gas species of one 
sign $q=\pm 1$, $n_q({\bf r}) \equiv n_q$ with ${\bf r}\in R^2$, 
is defined standardly as the thermal average 
$n_q = \langle \sum_j \delta_{q,q_j} \delta({\bf r}-{\bf r}_j)
\rangle_{\beta}$.
The charge neutrality of the system implies that $n_+ = n_- = n/2$,
where $n$ denotes the total number density of particles.
The species densities are expressible as field averages 
over the sine-Gordon action (\ref{2.2}) as follows
\begin{eqnarray} 
n_q & = & z_q \langle {\rm e}^{{\rm i} q b \phi} \rangle
\quad ( q= \pm 1 ) \nonumber \\
& = &  \frac{z}{2} \frac{\partial (-\omega)}{\partial z} . \label{2.9}
\end{eqnarray}
This equality and the relation (\ref{2.7}) determine explicitly 
the density-fugacity relationship \cite{Samaj00}:
\begin{equation} \label{2.10}
\frac{n^{1-(\beta/4)}}{z} = 2 
\left( \frac{\pi \beta}{8} \right)^{\beta/4}
\frac{\Gamma(1-(\beta/4))}{\Gamma(1+(\beta/4))}
\left[ F\left( \frac{1}{2}, \frac{\beta}{4-\beta}; 1+
\frac{\beta}{2(4-\beta)};1\right) \right]^{1-(\beta/4)} ,
\end{equation}
where $F\equiv {_2F_1}$ is the hypergeometric function.
Based on this density-fugacity relationship, the complete thermodynamics
of the 2D Coulomb gas was derived in the whole stability regime
of pointlike charges $\beta<2$ \cite{Samaj00}.

The excess (i.e., over ideal) chemical potential of the Coulomb-gas 
species $q=\pm 1$, $\mu_q^{\rm ex}$, is given by
\begin{equation} \label{2.11}
\exp\left( - \beta \mu_q^{\rm ex} \right) = \frac{n_q}{z_q}
= \langle {\rm e}^{{\rm i} q b \phi} \rangle ,
\quad q = \pm 1 . 
\end{equation}
Let $\mu_Q^{\rm ex}$ with arbitrarily valued real $Q$ represents an
extended definition of the excess chemical potential: $\mu_Q^{\rm ex}$
is the reversible work which has to be done in order to bring a pointlike
guest particle of charge $Q$ from infinity into the bulk interior
of the considered Coulomb gas.
It was shown in ref. \cite{Samaj05b} that $\mu_Q^{\rm ex}$ is expressible 
in the sine-Gordon format as follows
\begin{equation} \label{2.12}
\exp\left( -\beta \mu_Q^{\rm ex} \right) =
\langle {\rm e}^{{\rm i} Q b \phi} \rangle .
\end{equation}
When $Q=\pm 1$, one recovers the previous result (\ref{2.11}) 
valid for the Coulomb-gas constituents.
Due to the obvious symmetry relation
$\langle {\rm e}^{{\rm i}a\phi} \rangle = 
\langle {\rm e}^{-{\rm i}a\phi} \rangle$ valid for any real-valued
parameter $a$, it holds that $\mu_Q^{\rm ex} = \mu_{-Q}^{\rm ex}$.

A general formula for the expectation value of the exponential field
$\langle {\rm e}^{{\rm i}a\phi}\rangle$ was conjectured by 
Lukyanov and Zamolodchikov \cite{Lukyanov1}.
In the notation of Eq. (\ref{2.12}), $a = Q b$, their formula reads
\begin{equation} \label{2.13}
\langle {\rm e}^{{\rm i}Q b \phi} \rangle =
\left[ \frac{\pi z \Gamma(1-b^2)}{\Gamma(b^2)} \right]^{(Q b)^2/(1-b^2)} 
\exp\left[ I_b(Q) \right] , \qquad \vert Q \vert < \frac{1}{2 b^2}
\end{equation}
with
\begin{equation} \label{2.14}
I_b(Q) = \int_0^{\infty} \frac{{\rm d}t}{t} \left[
\frac{\sinh^2(2 Q b^2 t)}{2 \sinh(b^2 t) \sinh(t)
\cosh[(1-b^2)t]} - 2 Q^2 b^2 {\rm e}^{-2 t} \right] .
\end{equation} 
The integral (\ref{2.14}) is finite provided that 
$\vert Q\vert < 1/(2 b^2)$: at $\vert Q\vert = 1/(2 b^2)$
the integrated function behaves like $1/t$ for $t\to\infty$ what
causes the logarithmic divergence.
In the Coulomb-gas format, the interaction Boltzmann factor of
the guest $Q$ charge with an opposite unit plasma counterion
at distance $r$, $r^{-\beta\vert Q\vert}$, is integrable at small 2D
distances $r$ if and only if $\beta \vert Q\vert <2$.
In terms of the sine-Gordon coupling constant $b^2=\beta/4$,
the stability region for $\mu_Q^{\rm ex}$ is therefore expected
to be $\vert Q\vert < 1/(2 b^2)$ and the couple of Eqs. (\ref{2.13}) 
and (\ref{2.14}) passes the collapse test.

\subsection{Two-point densities}
At the two-particle statistical level, one introduces the two-body 
densities $n_{qq'}^{(2)}({\bf r},{\bf r}') = \langle \sum_{j\ne k}
\delta_{q,q_j} \delta({\bf r}-{\bf r}_j) 
\delta_{q',q_k} \delta({\bf r}'-{\bf r}_k) \rangle_{\beta}$
which are translationally invariant in the infinite 2D space,
$n_{qq'}^{(2)}({\bf r},{\bf r}') = 
n_{qq'}^{(2)}(\vert {\bf r}-{\bf r}'\vert).$ 
The two-body density is expressible as an average over the
sine-Gordon action (\ref{2.2}) as follows
\begin{equation} \label{2.15}
n_{qq'}^{(2)}(r) = z_q z_{q'} \langle
{\rm e}^{{\rm i}q b \phi({\bf 0})} {\rm e}^{{\rm i}q' b \phi({\bf r})}
\rangle ; \quad q,q' = \pm 1 .
\end{equation}
In close analogy with the previous case of one-body densities, 
it was shown in ref. \cite{Samaj05b} that the effective interaction energy 
of two guest charges immersed in the 2D Coulomb gas is expressible in terms
of more general two-point correlation functions of exponential fields:
$\langle {\rm e}^{{\rm i}Q b \phi({\bf 0})} {\rm e}^{{\rm i}Q' b \phi({\bf r})}
\rangle$ with real-valued charge parameters $Q$ and $Q'$.
A systematic generation of the short- and large-distance asymptotic
expansions for these two-point correlation functions is available
with the aid of special field-theoretical methods.

The short-distance expansion of $\langle {\rm e}^{{\rm i}Q b \phi({\bf 0})} 
{\rm e}^{{\rm i}Q' b \phi({\bf r})} \rangle$ can be obtained by using
the method of Operator Product Expansion (OPE) \cite{Wilson}. 
The OPE has the form \cite{Fateev1,Tellez05b}
\begin{equation} \label{2.16}
{\rm e}^{{\rm i}Q b \phi({\bf 0})} {\rm e}^{{\rm i}Q' b \phi({\bf r})} 
= \sum_{n=-\infty}^{\infty} \left\{ C_{QQ'}^{n,0}(r) 
{\rm e}^{{\rm i}(Q+Q'+n) b \phi}({\bf 0}) + \cdots \right\} ,
\end{equation}
where the dots stand for subleading contributions of the descendants
of ${\rm e}^{{\rm i}(Q+Q'+n)b\phi}$, like $(\partial \phi)^2 
({\bar\partial} \phi)^2 {\rm e}^{{\rm i}(Q+Q'+n)b\phi}$, etc. 
The coefficients $C$ read
\begin{equation} \label{2.17}
C_{QQ'}^{n,0}(r) = z^{\vert n\vert}
r^{4b^2[Q Q' + n(Q+Q') +n^2/2] + 2\vert n\vert (1-b^2)}
f_{QQ'}^{n,0}\left( z^2 r^{4-4b^2}\right) ,
\end{equation}
where the functions $f$ admit analytic series expansions
\begin{equation} \label{2.18}
f_{QQ'}^{n,0}(t) = \sum_{j=0}^{\infty} f_j^{n,0}(Q,Q') t^j .
\end{equation}
Each coefficient $f_j^{n,0}$ is expressible as a
$2(\vert n\vert + j)$-fold Coulomb integral defined in 
the infinite plane $R^2$.
The leading terms $f_0^{n,0}(Q,Q')$ in the series (\ref{2.18})
are expressible as 
\begin{eqnarray}
f_0^{0,0}(Q,Q') & = & 1 , \nonumber \\
f_0^{n,0}(Q,Q') & = & j_n(Q b^2,Q' b^2,b^2) \quad 
\mbox{for $n>0$,}  \label{2.19} \\
f_0^{n,0}(Q,Q') & = & j_{\vert n\vert}(-Q b^2,-Q' b^2,b^2) \quad 
\mbox{for $n<0$.} \nonumber
\end{eqnarray}
Here,
\begin{equation} \label{2.20}
j_n(a,a',b^2) = \frac{1}{n!} \int \prod_{j=1}^n \left( {\rm d}^2 r_j 
\vert {\bf r}_j\vert^{4a} \vert {\bf 1}-{\bf r}_j\vert^{4a'} \right)
\prod_{j<k} \vert {\bf r}_j-{\bf r}_k \vert^{4b^2}  
\end{equation}
with ${\bf 1}$ being a point on the unit circle, say $(1,0)$.
This integral is convergent if and only if the parameters $(a,a',b^2)$ 
fulfill the inequalities $a>-1/2$, $a'>-1/2$ and $a+a'<-(n-1)b^2-1/2$.
The integral (\ref{2.20}) was evaluated by Dotsenko and Fateev 
\cite{Dotsenko84,Dotsenko85}:
\begin{eqnarray} 
j_n(a,a',b^2) & = & \left[ \frac{\pi}{\gamma(b^2)} \right]^n
\prod_{j=1}^n \gamma(j b^2) \prod_{k=0}^{n-1} \gamma(1+2 a+k b^2)
\gamma(1+2 a' + k b^2) \nonumber \\ & & \times 
\gamma\left(-1-2 a -2 a' -(n-1+k) b^2\right) . \label{2.21}
\end{eqnarray}
Hereinafter, we use the notation $\gamma(t) = \Gamma(t)/\Gamma(1-t)$.
The result (\ref{2.21}) represents an analytic continuation of the
integral (\ref{2.20}) to all values of the parameters $(a,a',b^2)$.
We would like to emphasize that the OPE algebra (\ref{2.16}) is the
operation which can be used in any multi-point correlation function
of exponential fields to reduce its order as soon as a couple of
points is close to one another.

The large-distance asymptotic expansion of the truncated two-point
correlation functions 
\begin{equation} \label{2.22}
\langle {\rm e}^{{\rm i}Q b \phi({\bf 0})} 
{\rm e}^{{\rm i}Q' b \phi({\bf r})} \rangle^{\rm T} =
\langle {\rm e}^{{\rm i}Q b \phi({\bf 0})} 
{\rm e}^{{\rm i}Q' b \phi({\bf r})} \rangle -
\langle {\rm e}^{{\rm i}Q b \phi} \rangle
\langle {\rm e}^{{\rm i}Q' b \phi} \rangle
\end{equation}
can be obtained by using the form-factor method \cite{Smirnov}.
The form-factor representation is formally expressed as an infinite
convergent series over multi-particle intermediate states,
\begin{eqnarray}
\langle {\rm e}^{{\rm i}Q b \phi({\bf 0})} 
{\rm e}^{{\rm i}Q' b \phi({\bf r})} \rangle^{\rm T} & = &
\sum_{N=1}^{\infty} \frac{1}{N!} \sum_{\epsilon_1,\ldots,\epsilon_N}
\int_{-\infty}^{\infty} 
\frac{{\rm d}\theta_1 \ldots {\rm d}\theta_N}{(2\pi)^N} 
F_Q(\theta_1,\ldots,\theta_N)_{\epsilon_1,\ldots,\epsilon_N} \nonumber \\
& & \times \,
{^{\epsilon_N,\ldots,\epsilon_1}F}_{Q'}(\theta_N,\ldots,\theta_1) 
{\rm e}^{- r \sum_{j=1}^N m_{\epsilon_j} \cosh \theta_j} .
\label{2.23} 
\end{eqnarray} 
Here, $\epsilon$ indexes the particles of the sine-Gordon spectrum
(say $\epsilon=+/-$ for a soliton/antisoliton and $\epsilon=j$ for the
$B_j$ breather), the rapidity $\theta\in (-\infty,\infty)$ parameterizes 
the energy and the momentum of the particles, and $F$ denotes the
corresponding form factors.
In the large-distance limit $r\to\infty$, the dominant term on
the rhs of Eq. (\ref{2.23}) corresponds to the intermediate state
with the minimum value of the total particle mass 
$\sum_{j=1}^N m_{\epsilon_j}$, at the point of vanishing rapidities.
In the stability region of the pointlike Coulomb gas $0\le b^2 < 1/2$,
the lightest particle, which can exist in the spectrum alone, is
the $B_1$ breather of mass $m_1$.
For this particle, the one-particle form factors $F_Q(\theta)_1$
and ${^1F}_{Q'}(\theta) = F_{Q'}(\theta)_1$ were calculated in
refs. \cite{Lukyanov2} and \cite{Lukyanov3}:  
\begin{equation} \label{2.24}
F_Q(\theta)_1 = - {\rm i} \langle {\rm e}^{{\rm i}Q b\phi} \rangle
\sqrt{\pi\lambda} \frac{\sin(\pi\xi Q)}{\sin(\pi \xi)} ,
\end{equation}
where
\begin{equation} \label{2.25}
\lambda = \frac{4}{\pi} \sin(\pi \xi) \cos\left( \frac{\pi \xi}{2} \right)
\exp\left( - \int_0^{\pi\xi} \frac{{\rm d}t}{\pi} \frac{t}{\sin t} \right)
\end{equation}
and $\xi$ is defined in Eq. (\ref{2.4}).
Since the form factor (\ref{2.24}) does not depend on the rapidity, 
the integration over $\theta$ in (\ref{2.23}) can be done explicitly 
by using the relation
\begin{equation} \label{2.26}
\int_{-\infty}^{\infty} \frac{{\rm d}\theta}{2}
{\rm e}^{- r m_1 \cosh \theta} = K_0(m_1 r) \mathop{\sim}_{r\to\infty}
\left( \frac{\pi}{2 m_1 r} \right)^{1/2} {\rm e}^{-m_1 r} .
\end{equation}
Here, $K_0$ is the modified Bessel function of second kind.
Consequently,
\begin{equation} \label{2.27}
\frac{\langle {\rm e}^{{\rm i}Q b \phi({\bf 0})} 
{\rm e}^{{\rm i}Q' b \phi({\bf r})} \rangle^{\rm T}}{
\langle {\rm e}^{{\rm i}Q b \phi} \rangle
\langle {\rm e}^{{\rm i}Q' b \phi} \rangle}
\mathop{\sim}_{r\to\infty} - [ Q ] [ Q' ] \lambda 
\left( \frac{\pi}{2 m_1 r} \right)^{1/2} {\rm e}^{ - m_1 r} ,
\end{equation}
where the symbol $[ Q ]$ stands for the ratio
\begin{equation} \label{2.28}
[ Q ] = \frac{\sin( \pi \xi Q)}{\sin(\pi\xi)} .
\end{equation}
We see that, at large distance $r$, the truncated two-point correlation 
function factorizes into the product of separate phase contributions
$[Q]$ and $[Q']$.
The inverse correlation length in the exponential decay, $m_1$,
is determined exclusively by the Coulomb-gas system.
Using Eqs. (\ref{2.5}) -- (\ref{2.10}), $m_1$ is expressible as
\begin{equation} \label{2.29}
m_1 = \kappa \left[ \frac{\sin\left(\pi\beta/(4-\beta)\right)}{
\pi\beta/(4-\beta)} \right]^{1/2} .
\end{equation}
Here, $\kappa=\sqrt{2\pi\beta n}$ denotes the inverse Debye length;
in the high-temperature $\beta\to 0$ limit one has $m_1\sim\kappa$.
The formula (\ref{2.29}) describes the renormalization of the inverse
correlation length at a finite temperature.

\renewcommand{\theequation}{3.\arabic{equation}}
\setcounter{equation}{0}

\section{Chemical potential of the guest charge}

\subsection{Pointlike guest charge}
The chemical potential of the pointlike, without any loss of generality
say positive, guest charge $Q>0$, immersed in the bulk of the 2D Coulomb gas, 
is given by the relation (\ref{2.12}), complemented by the explicit formulas 
(\ref{2.13}) and (\ref{2.14}) for the expectation values of 
exponential fields in the sine-Gordon model.
Eqs. (\ref{2.13}) and (\ref{2.14}) are valid provided that $Q<Q_c$, where
\begin{equation} \label{3.1}
Q_c = \frac{1}{2 b^2} \quad  \left( = \frac{2}{\beta} \right)
\end{equation} 
is the ``collapse'' value of the guest charge.
Note that $Q_c>1$ in the considered stability regime of the Coulomb gas 
$\beta<2$ $(b^2<1/2)$.  

Eqs. (\ref{2.13}) and (\ref{2.14}) can be continued analytically
to the region $Q>Q_c$ by using the reflection method 
developed in refs. \cite{Fateev2,Fateev3}.   
Exploring a close relationship between the Liouville and sine-Gordon
theories, it was argued that the sine-Gordon expectation values
$\langle {\rm e}^{{\rm i}a\phi}\rangle$ obey the relation
\begin{equation} \label{3.2}
\langle {\rm e}^{{\rm i}a\phi} \rangle 
= R(a) \langle {\rm e}^{-{\rm i}(a+{\cal G})\phi} \rangle ,
\end{equation}
where ${\cal G} = b^{-1}-b$, and the reflection amplitude reads
\begin{equation} \label{3.3}
R(a) = \left[ \frac{\pi z \Gamma(1-b^2)}{b^2 \Gamma(1+b^2)}
\right]^{-(2 a +{\cal G})/b}
\frac{\Gamma\left( 1 + \frac{2a}{b} + \frac{\cal G}{b} \right)
\Gamma(1-2ab-{\cal G}b)}{\Gamma\left( 1 -\frac{2a}{b} - \frac{\cal G}{b} 
\right) \Gamma(1+2ab+{\cal G}b)} .
\end{equation}
Eqs. (\ref{2.13}) and (\ref{2.14}), in the notation of $a=Q b$ 
restricted to $\vert a\vert < 1/(2b)$, represent just the ``minimal'' 
solution of the reflection relations (\ref{3.2}) and (\ref{3.3}).
Using successively the reflection formula (\ref{3.2}), any expectation
value $\langle {\rm e}^{{\rm i}a\phi} \rangle$ with $\vert a\vert>1/(2b)$
can be reduced to a product of amplitudes of type (\ref{3.3}) and
the expectation value of the exponential field with a phase whose
absolute value $<1/(2b)$.

For our purpose, we take advantage of the symmetry property
$\langle {\rm e}^{{\rm i}a\phi} \rangle = 
\langle {\rm e}^{-{\rm i}a\phi} \rangle$ and rewrite the reflection
relation (\ref{3.2}), taken with $a=Qb$, in the following way
\begin{eqnarray}
\langle {\rm e}^{{\rm i}Qb\phi} \rangle & = &
\langle {\rm e}^{-{\rm i}Qb\phi} \rangle \nonumber \\
& = & R(-Qb) \langle {\rm e}^{{\rm i}(Qb-{\cal G})\phi} \rangle, \label{3.4}
\end{eqnarray}
where
\begin{equation} \label{3.5}
R(-Qb) = \left[ \frac{\pi z \Gamma(1-b^2)}{b^2 \Gamma(1+b^2)}
\right]^{2Q+1-b^{-2}}
\frac{\Gamma(-2Q+b^{-2})\Gamma(2 Q b^2 +b^2)}{\Gamma(2+2Q-b^{-2})
\Gamma(2-2 Q b^2 -b^2)} .
\end{equation}
While the minimal relations (\ref{2.13}) and (\ref{2.14}) are applicable to
$-Q_c<Q<Q_c$, the ``first-order'' relations (\ref{3.4}) and (\ref{3.5})
hold for $-Q_c<Q-({\cal G}/b)<Q_c$, which is equivalent to
\begin{equation} \label{3.6}
Q_c - 1 < Q < Q_c + (b^{-2}-1) .
\end{equation}
In the considered stability region $b^2<1/2$, one has 
$Q_c>1$, so that the lower bound for the guest charge $Q$ 
in Eq. (\ref{3.6}) is a positive number, and $b^{-2}-1>1$, so that 
the upper bound for $Q$ in Eq. (\ref{3.6}) is larger than $Q_c+1$.

The only singularities (in particular, simple poles) of the reflection
amplitude $R(-Qb)$ in Eq. (\ref{3.5}) come from the Gamma function 
$\Gamma(-2Q+b^{-2})$, when its argument attains zero or a negative integer, 
i.e., at the infinite sequence of equidistant $Q$-points
\begin{equation} \label{3.7}
Q_n^* = Q_c + \frac{n}{2}; \quad n=0,1,2,\ldots .
\end{equation}
The first three singular $Q$-points, which will be of special interest in
what follows, explicitly read
\begin{equation} \label{3.8}
Q_0^* = Q_c, \quad Q_1^* = Q_c + \frac{1}{2}, \quad 
Q_2^* = Q_c + 1 ;
\end{equation}
note that these points lie in the range (\ref{3.6}) of the applicability
of the first-order reflection relations (\ref{3.4}) and (\ref{3.5}).
Using the residuum formula for the Gamma function
\begin{equation} \label{3.9}
\lim_{\epsilon\to 0^+} \Gamma(-n+\epsilon) =
\frac{(-1)^n}{n!} \frac{1}{\epsilon} \quad
(n=0,1,2,\ldots) ,
\end{equation}
Eqs. (\ref{3.4}) and (\ref{3.5}) imply, as $\epsilon\to 0^+$, 
the following singular behaviors
\begin{eqnarray}
\langle {\rm e}^{{\rm i}Qb\phi} \rangle & \sim & \frac{\pi z}{2 b^2 \epsilon} 
\left\langle {\rm e}^{{\rm i}( Q_0^* - 1 ) b \phi} \right\rangle
\quad \mbox{for $Q=Q_0^*-\epsilon$,} \label{3.10} \\
\langle {\rm e}^{{\rm i}Qb\phi} \rangle & \sim & 
- \frac{1}{4\epsilon} \left[ \frac{\pi z \Gamma(1-b^2)}{b^2 \Gamma(1+b^2)}
\right]^2 \frac{\Gamma(1+2b^2)}{\Gamma(1-2b^2)} \nonumber \\
& & \times \left\langle {\rm e}^{{\rm i}( Q_1^* - 2 ) b \phi} \right\rangle 
\quad \mbox{for $Q=Q_1^*-\epsilon$,} \label{3.11}
\end{eqnarray}
and so on.

\subsection{Guest charge with hard core}
Let us introduce an impenetrable hard disc of radius $\sigma$ around 
the positively charged $Q>0$ guest particle localized in the bulk interior 
of the 2D Coulomb gas, say at the origin ${\bf r}={\bf 0}$.
The radius $\sigma$ has the dimension of length; the most natural
dimensionless quantity involving $\sigma$ is chosen as
\begin{equation} \label{3.12}
{\hat \sigma} = m_1 \sigma ,
\end{equation}
where $m_1$ is the inverse correlation length of the Coulomb-gas 
constituents given by Eq. (\ref{2.29}).
The corresponding excess chemical potential of the guest charge
will be denoted by $\mu_Q^{\rm ex}(\sigma)$.
As soon as $\sigma>0$, $\mu_Q^{\rm ex}(\sigma)$ must be finite for any
value of $Q$.
It was shown in ref. \cite{Samaj05b} that $\mu_Q^{\rm ex}(\sigma)$
is expressible in the sine-Gordon format as follows
\begin{equation} \label{3.13}
\exp\left[ -\beta \mu_Q^{\rm ex}(\sigma) \right] =
\langle {\rm e}^{{\rm i}Qb\phi({\bf 0})} \rangle_{\sigma} ,
\end{equation}
where the average $\langle \cdots \rangle_{\sigma}$ is defined by
\begin{eqnarray}
\langle \cdots \rangle_{\sigma} & = & \frac{1}{\Xi(z)}
\int {\cal D}\phi\, {\rm e}^{-S_{\sigma}(z)} \cdots , \label{3.14} \\
S_{\sigma}(z) & = & S(z) + z \int_{r<\sigma} {\rm d}^2 r
\left[ {\rm e}^{{\rm i}b\phi({\bf r})} + {\rm e}^{-{\rm i}b\phi({\bf r})}
\right] . \label{3.15}
\end{eqnarray}
Here, $S(z)$ is the usual sine-Gordon action (\ref{2.2}).

The averaged quantity 
$\langle {\rm e}^{{\rm i}Qb\phi({\bf 0})} \rangle_{\sigma}$
can be formally expanded around $S(z)$:
\begin{equation} \label{3.16}
\langle {\rm e}^{{\rm i}Qb\phi({\bf 0})} \rangle_{\sigma} 
= \langle {\rm e}^{{\rm i}Qb\phi} \rangle +
\sum_{n=1}^{\infty} \frac{(-z)^n}{n!} I_Q^{(n)}(\sigma) ,
\end{equation}
where
\begin{eqnarray}
I_Q^{(1)}(\sigma) & = & \int_{r<\sigma} {\rm d}^2 r \left[ 
\langle {\rm e}^{{\rm i}Qb\phi({\bf 0})} 
{\rm e}^{{\rm i}b\phi({\bf r})} \rangle
+ \langle {\rm e}^{{\rm i}Qb\phi({\bf 0})} 
{\rm e}^{-{\rm i}b\phi({\bf r})} \rangle \right] , \label{3.17} \\
I_Q^{(2)}(\sigma) & = & \int_{r<\sigma} {\rm d}^2 r 
\int_{r'<\sigma} {\rm d}^2 r' \left[ 
\langle {\rm e}^{{\rm i}Qb\phi({\bf 0})} {\rm e}^{{\rm i}b\phi({\bf r})} 
{\rm e}^{{\rm i}b\phi({\bf r}')} \rangle +
\langle {\rm e}^{{\rm i}Qb\phi({\bf 0})} {\rm e}^{{\rm i}b\phi({\bf r})} 
{\rm e}^{-{\rm i}b\phi({\bf r}')} \rangle \right. \nonumber\\
& & \left. + \langle {\rm e}^{{\rm i}Qb\phi({\bf 0})} 
{\rm e}^{-{\rm i}b\phi({\bf r})} 
{\rm e}^{{\rm i}b\phi({\bf r}')} \rangle +
\langle {\rm e}^{{\rm i}Qb\phi({\bf 0})} {\rm e}^{-{\rm i}b\phi({\bf r})} 
{\rm e}^{-{\rm i}b\phi({\bf r}')} \rangle \right] , \label{3.18}
\end{eqnarray}
and so on.
The expansion (\ref{3.16}) represents the basis for a systematic generation 
of ${\hat\sigma}$-corrections to the pure sine-Gordon expectation value
$\langle {\rm e}^{{\rm i}Qb\phi}\rangle$, in the limit ${\hat\sigma}\to 0$.  
To be more particular, for a given value of the guest charge $Q$, 
the rhs of Eq. (\ref{3.16}) contains ${\hat\sigma}$-dependent terms 
which either vanish (and so they are irrelevant and omitted) or diverge 
(and so they are relevant and preserved) in the considered limit 
${\hat\sigma}\to 0$.  
Since the chemical potential $\mu_Q^{\rm ex}(\sigma>0)$ in Eq. (\ref{3.13})
is finite for any $Q$, these relevant ${\hat\sigma}$-dependent terms 
must cancel the singularities of $\langle {\rm e}^{{\rm i}Qb\phi}\rangle$ 
occurring at the points $\{ Q_n^* \}_{n=0}^{\infty}$ given by Eq. (\ref{3.7}).
The main advantage of the expansion (\ref{3.16}) is that, as $Q$ increases
from 0 to $\infty$, the singularities of 
$\langle {\rm e}^{{\rm i}Qb\phi}\rangle$ at the points
$Q_0^*, Q_1^*,\ldots$ are eliminated successively via the respective
integral terms $I_Q^{(1)}, I_Q^{(2)}, \ldots$, where the multi-point
correlation functions are evaluated by using the short-distance OPE 
scheme described by Eqs. (\ref{2.16})--(\ref{2.21}).
We shall document this claim on lower levels in the next paragraphs. 

In the stability region of the pointlike guest charge $0\le Q<Q_0^*$, one has
\begin{equation} \label{3.19}
\langle {\rm e}^{{\rm i}Qb\phi({\bf 0})} \rangle_{\sigma} 
\sim  \langle {\rm e}^{{\rm i}Qb\phi} \rangle
\quad \mbox{as ${\hat\sigma}\to 0$}
\end{equation}
in the sense that all ${\hat\sigma}$-dependent terms on the rhs of
Eq. (\ref{3.16}) vanish in the limit ${\hat\sigma}\to 0$.
This is a direct consequence of the fact that the chemical potential
$\mu_Q^{\rm ex}(\sigma)$, given by Eq. (\ref{3.13}), is finite even for 
${\hat\sigma}=0$ in the guest-charge stability region $\vert Q\vert <Q_0^*$.

When $Q_0^*-\epsilon\le Q < Q_1^*$ (in what follows, under $\epsilon$
we shall understand a positive number going to zero, $\epsilon\to 0^+$), 
the first term on the rhs of Eq. (\ref{3.16}) has to be taken into account.
The corresponding integral $I_Q^{(1)}(\sigma)$, defined in Eq. (\ref{3.17}),
consists of two parts: the first term corresponds to the correlation function
of the pointlike guest charge $Q$ with one coion (unit Coulomb-gas charge
of the same sign as $Q$), the second term corresponds to the correlation
function of the $Q$-charge with one counterion.
In the considered limit ${\hat\sigma}\to 0$, the systematic
short-distance expansions of the underlying correlation functions,
based on the OPE scheme (\ref{2.16})--(\ref{2.21}), has to be applied.
The dominant contribution comes from the correlation function of the
$Q$ charge and one counterion, in the leading short-distance order
corresponding to the bare-Coulomb Boltzmann factor:
\begin{equation} \label{3.20}
\langle {\rm e}^{{\rm i}Qb\phi({\bf 0})} {\rm e}^{-{\rm i}b\phi({\bf r})} 
\rangle \mathop{\sim}_{r\to 0}   
\langle {\rm e}^{{\rm i}(Q-1)b\phi} \rangle r^{-4Qb^2}.
\end{equation}
Consequently,
\begin{equation} \label{3.21}
I_Q^{(1)}(\sigma) \sim 2\pi
\frac{\langle {\rm e}^{{\rm i}(Q-1)b\phi} \rangle}{m_1^{2-4Qb^2}}
\frac{{\hat\sigma}^{2-4Qb^2}}{2-4Qb^2} \quad \mbox{as ${\hat\sigma}\to 0$.}
\end{equation}  
After simple algebraic manipulations, Eq. (\ref{3.16}) then reads
\begin{equation} \label{3.22}
\langle {\rm e}^{{\rm i}Qb\phi({\bf 0})} \rangle_{\sigma}
\mathop{\sim}_{{\hat\sigma}\to 0} 
\langle {\rm e}^{{\rm i}Qb\phi} \rangle - \frac{\pi z}{2 b^2} 
\frac{\langle {\rm e}^{{\rm i}(Q-1)b\phi} \rangle}{m_1^{4b^2(Q_0^*-Q)}}
\frac{{\hat\sigma}^{4b^2(Q_0^*-Q)}}{Q_0^*-Q} .
\end{equation}
Close to the collapse value of the guest charge, i.e., 
when $Q=Q_0^*-\epsilon$, one expands
\begin{equation} \label{3.23}
\frac{{\hat\sigma}^{4b^2(Q_0^*-Q)}}{Q_0^*-Q} 
= \frac{1}{\epsilon} + 4 b^2 \ln {\hat\sigma} + O(\epsilon) .
\end{equation}
Inserting this expansion into Eq. (\ref{3.22}), the leading singular
term of order $\epsilon^{-1}$ cancels exactly with its counterpart in
$\langle{\rm e}^{{\rm i}Qb\phi}\rangle$ [see Eq. (\ref{3.10})],
and one ends up with the logarithmic dependence on the short-distance
cutoff ${\hat\sigma}$:
\begin{equation} \label{3.24}
\langle {\rm e}^{{\rm i}Q_0^* b\phi({\bf 0})} \rangle_{\sigma}
\mathop{\sim}_{{\hat\sigma}\to 0} - 2 \pi z 
\langle {\rm e}^{{\rm i}(Q_0^*-1)b\phi} \rangle \ln {\hat\sigma} .  
\end{equation}
The outlined cancellation of singularities at $Q_0^*$, due to
the presence of the hard core around the guest charge, is an important 
check of the consistency of the present expansion procedure.
For $Q>Q_0^*$, the exponent of ${\hat\sigma}$ in Eq. (\ref{3.22})
is negative as was anticipated.

When $Q_1^*-\epsilon\le Q<Q_2^*$, in the limit ${\hat\sigma}\to 0$,
two additional relevant ${\hat\sigma}$-contributions arise on the rhs of 
Eq. (\ref{3.16}).
The first contribution has the origin, as in the above paragraph,
in the integral $I_Q^{(1)}(\sigma)$ (\ref{3.17}) as the result of
the next-to-leading expansion term for the correlation function
of the $Q$ charge with one counterion:
\begin{equation} \label{3.25}
\langle {\rm e}^{{\rm i}Qb\phi({\bf 0})} {\rm e}^{-{\rm i}b\phi({\bf r})} 
\rangle \mathop{\sim}_{r\to 0}   
\langle {\rm e}^{{\rm i}(Q-1)b\phi} \rangle r^{-4Qb^2}
+ z\, j_1(-Qb^2,b^2,b^2) \langle {\rm e}^{{\rm i}(Q-2)b\phi} \rangle 
r^{2+4b^2-8Qb^2} .
\end{equation}
This expansion was derived from the OPE scheme (\ref{2.16})--(\ref{2.21})
assuming that $Q>1$, which is indeed true for the considered range of
$Q$ values.
Thus,
\begin{eqnarray}
I_Q^{(1)}(\sigma) & \sim & 2\pi 
\frac{\langle {\rm e}^{{\rm i}(Q-1)b\phi} \rangle}{m_1^{2-4Qb^2}}
\frac{{\hat\sigma}^{2-4Qb^2}}{2-4Qb^2} 
+ 2\pi z\, j_1(-Qb^2,b^2,b^2) \nonumber \\
& & \times
\frac{\langle {\rm e}^{{\rm i}(Q-2)b\phi} \rangle}{m_1^{4+4b^2-8Qb^2}}
\frac{{\hat\sigma}^{4+4b^2-8Qb^2}}{4+4b^2-8Qb^2} 
\quad \mbox{as ${\hat\sigma}\to 0$.} \label{3.26}
\end{eqnarray}
According to the Dotsenko-Fateev result (\ref{2.21}), it holds
\begin{equation} \label{3.27}
j_1(-Qb^2,b^2,b^2) = \pi \gamma(1-2Qb^2) \gamma(1+2b^2)
\gamma(-1-2b^2+2Qb^2) .
\end{equation}
The second relevant ${\hat\sigma}$-contribution has the origin in
the integral $I_Q^{(2)}(\sigma)$ (\ref{3.18}), namely, in the
correlation function
$\langle {\rm e}^{{\rm i}Qb\phi({\bf 0})} {\rm e}^{-{\rm i}b\phi({\bf r})} 
{\rm e}^{-{\rm i}b\phi({\bf r}')} \rangle$  
of the $Q$ charge with two electrolyte counterions.
At short distances $r$ and $r'$, this three-point correlation function 
is governed by the bare-Coulomb Boltzmann factor and behaves like
\begin{equation} \label{3.28}
\langle {\rm e}^{{\rm i}Qb\phi({\bf 0})} {\rm e}^{-{\rm i}b\phi({\bf r})} 
{\rm e}^{-{\rm i}b\phi({\bf r}')} \rangle \mathop{\sim}_{r,r'\to 0}
\langle {\rm e}^{{\rm i}(Q-2)b\phi} \rangle r^{-4 Q b^2}
(r')^{-4 Q b^2} \vert {\bf r}-{\bf r}'\vert^{4b^2} .
\end{equation}
Thus,
\begin{equation} \label{3.29}
I_Q^{(2)}(\sigma) \sim J_Q^{(2)} 
\frac{\langle {\rm e}^{{\rm i}(Q-2)b\phi} \rangle}{m_1^{4+4b^2-8Qb^2}}
{\hat\sigma}^{4+4b^2-8Qb^2} \quad \mbox{as ${\hat\sigma}\to 0$,}
\end{equation}
where $J_Q^{(2)}$ is the Coulomb integral defined inside the unit disc:
\begin{equation} \label{3.30}
J_Q^{(2)} = \int_{r<1} {\rm d}^2 r \int_{r'<1} {\rm d}^2 r'\,
r^{-4 Q b^2} (r')^{-4 Q b^2} \vert {\bf r}-{\bf r}'\vert^{4b^2} .
\end{equation}
The derivation of its series representation
\begin{equation} \label{3.31}
J_Q^{(2)} = \frac{\pi^2}{1+b^2-2Qb^2} \sum_{j=0}^{\infty} \frac{1}{1-2Qb^2+j} 
\left[ \frac{\Gamma(j-2b^2)}{j!\Gamma(-2b^2)} \right]^2
\end{equation}
is outlined in the Appendix, see the final Eq. (\ref{A.7}).
Inserting the formula (\ref{3.26}) together with Eqs. (\ref{3.29}) and 
(\ref{3.31}) into the basic series expansion (\ref{3.16}), 
after simple algebra one finally arrives at
\begin{eqnarray} 
\langle {\rm e}^{{\rm i}Qb\phi({\bf 0})} \rangle_{\sigma}
& \displaystyle{\mathop{\sim}_{{\hat\sigma}\to 0}} & 
\langle {\rm e}^{{\rm i}Qb\phi} \rangle - \frac{\pi z}{2 b^2} 
\frac{\langle {\rm e}^{{\rm i}(Q-1)b\phi} \rangle}{m_1^{4b^2(Q_0^*-Q)}}
\frac{{\hat\sigma}^{4b^2(Q_0^*-Q)}}{Q_0^*-Q}  \nonumber \\
& & + f_Q^{(2)} \frac{(\pi z)^2}{4 b^2} 
\frac{\langle {\rm e}^{{\rm i}(Q-2)b\phi} \rangle}{m_1^{8b^2(Q_1^*-Q)}}
\frac{{\hat\sigma}^{8b^2(Q_1^*-Q)}}{Q_1^*-Q} , \label{3.32}
\end{eqnarray}
where
\begin{equation} \label{3.33} 
f_Q^{(2)}  = \sum_{j=0}^{\infty} \frac{1}{1-2Qb^2+j} 
\left[ \frac{\Gamma(j-2b^2)}{j!\Gamma(-2b^2)} \right]^2
- \frac{1}{\pi} j_1(-Qb^2,b^2,b^2) .
\end{equation}
In addition to the previous result (\ref{3.22}), Eq. (\ref{3.32})
contains the ${\hat\sigma}$-dependent term whose exponent is negative,
and therefore relevant, just for $Q>Q_1^*$.
This term removes the singularity of the pointlike expectation value
$\langle {\rm e}^{{\rm i}Qb\phi} \rangle$ at $Q=Q_1^*$ 
described by the relation (\ref{3.11}).
To show this fact, we first recall the formula (\ref{A.11}) derived
in the Appendix:
\begin{equation} \label{3.34}
\sum_{j=0}^{\infty} \frac{1}{1-2Q_1^*b^2+j} 
\left[ \frac{\Gamma(j-2b^2)}{j!\Gamma(-2b^2)} \right]^2
= \frac{1}{2\pi} j_1(-Q_1^*b^2,b^2,b^2) .
\end{equation}
Thus, at $Q=Q_1^*$, Eq. (\ref{3.33}) yields 
\begin{eqnarray} 
f_{Q_1^*}^{(2)} & = & - \frac{1}{2\pi} j_1(-Q_1^*b^2,b^2,b^2) 
= - \frac{1}{2} \left[ \gamma(-b^2) \right]^2 \gamma(1+2b^2)
\nonumber \\ & = & 
\frac{1}{b^2} \left[ \frac{\Gamma(1-b^2)}{\Gamma(1+b^2)} \right]^2
\frac{\Gamma(1+2b^2)}{\Gamma(1-2b^2)} . \label{3.35}
\end{eqnarray}
Here, we have used the Dotsenko-Fateev result (\ref{3.27}) and
the standard relation $\Gamma(x+1) = x \Gamma(x)$ for the Gamma functions.
When $Q=Q_1^*-\epsilon$, after inserting the expansion
\begin{equation} \label{3.36}
\frac{{\hat\sigma}^{8b^2(Q_1^*-Q)}}{Q_1^*-Q} 
= \frac{1}{\epsilon} + 8 b^2 \ln {\hat\sigma} + O(\epsilon) 
\end{equation}
and the explicit form of $f_{Q_1^*}$ (\ref{3.35}) into Eq. (\ref{3.32}),
the leading singular term of order $\epsilon^{-1}$ cancels exactly with
its counterpart in $\langle {\rm e}^{{\rm i}Qb\phi} \rangle$ [see 
Eq. (\ref{3.11})]. 
As before at the collapse value $Q_0^*$ [see Eq. (\ref{3.24})], 
one ends up with a logarithmic dependence on the short-distance cutoff 
${\hat\sigma}$ at the point $Q_1^*$.

Note that the exponents of the two ${\hat\sigma}$-dependent terms
on the rhs of Eq. (\ref{3.32}) satisfy the inequality
$4b^2(Q_0^*-Q)<8b^2(Q_1^*-Q)$ up to $Q=2Q_1^*-Q_0^*=Q_2^*$.
This means that, in the ${\hat\sigma}\to 0$ limit, the first
${\hat\sigma}$-dependent term with the exponent $4b^2(Q_0^*-Q)$
is dominant in the whole interval $Q_0^*\le Q<Q_2^*$:
\begin{equation} \label{3.37}
\langle {\rm e}^{{\rm i}Qb\phi({\bf 0})} \rangle_{\sigma}
\mathop{\sim}_{{\hat\sigma}\to 0} \frac{\pi z}{2 b^2} 
\frac{\langle {\rm e}^{{\rm i}(Q-1)b\phi} \rangle}{m_1^{4b^2(Q_0^*-Q)}}
\frac{{\hat\sigma}^{4b^2(Q_0^*-Q)}}{Q-Q_0^*} , \quad
Q_0^* \le Q < Q_2^* .
\end{equation}
Crossing the next singular point $Q_2^*$, this term becomes subleading 
and the second ${\hat\sigma}$-dependent term with the exponent 
$8b^2(Q_1^*-Q)$ takes the dominant role, it turns out that up to $Q_4^*$.

One can proceed along the above lines to higher-order 
${\hat\sigma}$-contributions, with an increasing amount of
algebraic laboriousness.
In this paragraph, we indicate the general structure of the
relevant ${\hat\sigma}$-contributions on the rhs of Eq. (\ref{3.16}).
From among correlation functions in the integral $I_Q^{(n)}(\sigma)$,
the important one corresponds to the configuration of one guest charge $Q$ 
surrounded by $n$ counterions,
$\langle {\rm e}^{{\rm i}Qb\phi({\bf 0})} {\rm e}^{-{\rm i}b\phi({\bf r}_1)}
\ldots {\rm e}^{-{\rm i}b\phi({\bf r}_n)} \rangle$.
The leading short-distance expansion of this correlation function
is governed by the bare-Coulomb Boltzmann factor:
\begin{equation} \label{3.38}
\langle {\rm e}^{{\rm i}Qb\phi({\bf 0})} {\rm e}^{-{\rm i}b\phi({\bf r}_1)}
\cdots {\rm e}^{-{\rm i}b\phi({\bf r}_n)} \rangle
\mathop{\sim}_{r_1,\ldots,r_n\to 0} 
\langle {\rm e}^{{\rm i}(Q-n)b\phi} \rangle
r_1^{-4Qb^2} \cdots r_n^{-4Qb^2} 
\prod_{i<j} \vert {\bf r}_i-{\bf r}_j \vert^{4b^2} .
\end{equation}
Rescaling the 2D spatial coordinates ${\bf r}_1,\ldots,{\bf r}_n$ by
$\sigma$ and regarding the equality 
$2n-4nQb^2+2n(n-1)b^2 = 4nb^2(Q_{n-1}^*-Q)$, one obtains
\begin{equation} \label{3.39}
I_Q^{(n)}(\sigma) \sim J_Q^{(n)}
\frac{\langle {\rm e}^{{\rm i}(Q-n)b\phi} \rangle}{m_1^{4nb^2(Q_{n-1}^*-Q)}}
{\hat\sigma}^{4nb^2(Q_{n-1}^*-Q)} \quad \mbox{as ${\hat\sigma}\to 0$} ,
\end{equation}
where $J_Q^{(n)}$ is a $2n$-fold Coulomb integral defined in the unit disc.
This ${\hat\sigma}$-contribution has its counterparts, which possess
the structure of Eq. (\ref{3.39}) except of prefactors different from
$J_Q^{(n)}$, in each of the lower-order integrals $I_Q^{(1)}(\sigma),\ldots,
I_Q^{(n-1)}(\sigma)$ in the series (\ref{3.16}) due to the existence of
higher-order terms of the short-distance expansions of 
the correlation functions under integration.
In analogy with Eq. (\ref{3.32}), the summation over all 
${\hat\sigma}$-contributions belonging to the family (\ref{3.39})
can be formally represented as
\begin{equation} \label{3.40}
f_Q^{(n)} \frac{(-2\pi z)^n}{n! (4 n b^2)} 
\frac{\langle {\rm e}^{{\rm i}(Q-n)b\phi} \rangle}{m_1^{4nb^2(Q_{n-1}^*-Q)}}
\frac{{\hat\sigma}^{4nb^2(Q_{n-1}^*-Q)}}{Q_{n-1}^*-Q} .
\end{equation}
The term (\ref{3.40}) becomes relevant for $Q>Q_{n-1}^*$; 
the prefactor $f_Q^{(n)}$ is such that it cancels
exactly the singularity of the pointlike mean value 
$\langle {\rm e}^{{\rm i}Qb\phi} \rangle$ at $Q_{n-1}^*$.
Two consecutive $n$th and $(n+1)$st ${\hat\sigma}$-contributions of 
type (\ref{3.40}) interchange their dominant role when the respective
exponents $4nb^2(Q_{n-1}^*-Q)$ and $4(n+1)b^2(Q_n^*-Q)$ coincide,
i.e., at the point $Q = (n+1)Q_n^*-nQ_{n-1}^* = Q_{2n}^*$.
This means that, in the ${\hat\sigma}\to 0$ limit,
$\langle {\rm e}^{{\rm i}Qb\phi({\bf 0})} \rangle_{\sigma}$
is dominated by the $n$th ${\hat\sigma}$-contribution (\ref{3.40})
in the range $Q_{2(n-1)}^* \le Q < Q_{2n}^*$.

\renewcommand{\theequation}{4.\arabic{equation}}
\setcounter{equation}{0}

\section{Renormalized charge}

\subsection{Pointlike guest charge}
The pointlike guest particle of charge $Q>0$, localized at the origin ${\bf 0}$
and surrounded by the infinite 2D Coulomb gas, evokes position-dependent
density profiles $n_q({\bf r})$ of the electrolyte species $q=\pm 1$.
In the sine-Gordon field representation, these density profiles
are given by \cite{Samaj05b} 
\begin{equation} \label{4.1}
n_q({\bf r}) = z \frac{\langle {\rm e}^{{\rm i}Qb\phi({\bf 0})}
{\rm e}^{{\rm i}qb\phi({\bf r})} \rangle}{\langle 
{\rm e}^{{\rm i}Qb\phi} \rangle} , \quad q=\pm 1 .
\end{equation}
At large distances from the guest charge, the form-factor
result (\ref{2.27}) implies
\begin{equation} \label{4.2}
n_q(r) - n_q \sim - n_q [q] [Q] \lambda 
\left( \frac{\pi}{2 m_1 r} \right)^{1/2} \exp(-m_1 r) 
\quad \mbox{as $r\to\infty$.}
\end{equation}
Here, $[q]=q$ for $q=\pm 1$, the function $[Q]$ (\ref{2.28}) is 
expressible in terms of the inverse temperature $\beta$ as follows
\begin{equation} \label{4.3}
[Q] = \frac{\sin[\pi\beta Q/(4-\beta)]}{\sin[\pi\beta/(4-\beta)]} ,
\end{equation}
and the parameter $\lambda$ (\ref{2.25}) as follows
\begin{equation} \label{4.4}
\lambda = \frac{4}{\pi} \sin\left( \frac{\pi\beta}{4-\beta} \right)
\cos\left( \frac{\pi\beta}{2(4-\beta)} \right)
\exp\left\{ - \int_0^{\pi\beta/(4-\beta)} \frac{{\rm d}t}{\pi}
\frac{t}{\sin t} \right\} .
\end{equation}

Using Eq. (\ref{4.2}), the induced charge density $\rho$ of electrolyte 
particles, defined by $\rho({\bf r}) = n_+({\bf r}) - n_-({\bf r})$, reads 
\begin{equation} \label{4.5}
\rho(r) \mathop{\sim}_{r\to\infty} - n [Q] \lambda
\left( \frac{\pi}{2 m_1 r} \right)^{1/2} \exp(-m_1 r) .
\end{equation}
The average electrostatic potential $\psi$ is related to
the charge-density profile through the 2D Poisson equation,
\begin{equation} \label{4.6}
\Delta \psi({\bf r}) = - 2 \pi \rho({\bf r}) .
\end{equation}
Considering the circularly symmetric Laplacian 
$\Delta = r^{-1} {\rm d}_r(r{\rm d}_r)$, 
the asymptotic formula (\ref{4.5}) gives
\begin{equation} \label{4.7}
\beta \psi(r) \mathop{\sim}_{r\to\infty} [Q] \lambda 
\left( \frac{\kappa}{m_1} \right)^2  
\left( \frac{\pi}{2 m_1 r} \right)^{1/2} \exp(-m_1 r) .
\end{equation}
In the Debye-H\"uckel limit $\beta\to 0$, it holds $[Q]\sim Q$,
$m_1\sim \kappa$ and $\lambda\sim \beta$.
Eq. (\ref{4.7}) then reduces to 
\begin{equation} \label{4.8}
\beta \psi_{\rm DH}(r) \mathop{\sim}_{r\to\infty} 
\beta Q \left( \frac{\pi}{2 \kappa r} \right)^{1/2} \exp(-\kappa r) .
\end{equation}
Eq. (\ref{4.7}) is consistent with Eq. (\ref{4.8}), in terms of
the dimensionless combinations $m_1 r$ and $\kappa r$, respectively,
when one introduces the renormalized charge $Q_{\rm ren}$ as follows 
\cite{Samaj05b}
\begin{equation} \label{4.9}
\beta Q_{\rm ren} = [Q] \lambda \left( \frac{\kappa}{m_1} \right)^2 .
\end{equation} 

We see that the dependence of the renormalized charge $Q_{\rm ren}$ on the 
bare charge $Q$ enters only via the function $[Q]$ defined by Eq. (\ref{4.3}).
The rigorous validity of the result (\ref{4.9}) is restricted to 
the guest-charge stability region $0\le Q<Q_c$.
Within this range of $Q$ values, the function $[Q]$ exhibits the maximum 
at $Q=Q_c-1/2$.
It can be readily checked that at the collapse value of the guest
charge this function satisfies the important relation
\begin{equation} \label{4.10}
[Q_c] = [Q_c - 1]
\end{equation}
which was not noted in the previous work \cite{Samaj05b}.
The relation (\ref{4.10}) reflects the effect of the counterion 
condensation: the renormalized charge associated with the collapse value 
$Q_c$ of the bare charge is identical to that of the bare charge $Q_c-1$ 
because of the condensation of one counterion from the electrolyte onto 
the guest particle. 

\subsection{Guest charge with hard core}
We now consider the positive guest charge $Q$ possessing the hard core
of radius $\sigma$.
The evoked density profiles $n_q({\bf r})$ of the electrolyte species
$q=\pm 1$ vanish inside the hard disc, $n_q({\bf r}) = 0$
for $r\le \sigma$.
Using the formalism of ref. \cite{Samaj05b} it can be shown that
outside the hard-core region the number densities of the
electrolyte species are given by the following generalization of 
the relation (\ref{4.1})
\begin{equation} \label{4.11}
n_q({\bf r}) = z \frac{\langle {\rm e}^{{\rm i}Qb\phi({\bf 0})}
{\rm e}^{{\rm i}qb\phi({\bf r})} \rangle_{\sigma}}{\langle 
{\rm e}^{{\rm i}Qb\phi({\bf 0})} \rangle_{\sigma}} \quad
\mbox{for $r>\sigma$,}
\end{equation}
where the average $\langle \cdots \rangle_{\sigma}$ is defined in
Eqs. (\ref{3.14}) and (\ref{3.15}).
The relation (\ref{4.11}) can be rewritten in a more convenient form:
\begin{equation} \label{4.12}
n_q({\bf r}) - n_q = n_q \frac{\langle {\rm e}^{{\rm i}Qb\phi({\bf 0})}
{\rm e}^{{\rm i}qb\phi({\bf r})} \rangle_{\sigma} -
\langle {\rm e}^{{\rm i}Qb\phi({\bf 0})} \rangle_{\sigma}
\langle {\rm e}^{{\rm i}qb\phi} \rangle}{\langle 
{\rm e}^{{\rm i}Qb\phi({\bf 0})} \rangle_{\sigma}
\langle {\rm e}^{{\rm i}qb\phi} \rangle}, 
\quad r>\sigma .
\end{equation}

The one-point quantity 
$\langle {\rm e}^{{\rm i}Qb\phi({\bf 0})} \rangle_{\sigma}$ was analyzed 
in Section 3.2, the large-distance analysis of the 
two-point correlation functions $\langle {\rm e}^{{\rm i}Qb\phi({\bf 0})}
{\rm e}^{{\rm i}qb\phi({\bf r})} \rangle_{\sigma}$ $(q=\pm 1)$
will proceed analogously.
Expanding formally $\langle \cdots \rangle_{\sigma}$ around the
sine-Gordon action $S(z)$, one obtains
\begin{equation} \label{4.13}
\langle {\rm e}^{{\rm i}Qb\phi({\bf 0})}
{\rm e}^{{\rm i}qb\phi({\bf r})} \rangle_{\sigma} =
\langle {\rm e}^{{\rm i}Qb\phi({\bf 0})}
{\rm e}^{{\rm i}qb\phi({\bf r})} \rangle
+ \sum_{n=1}^{\infty} \frac{(-z)^n}{n!}
I_{Q,q}^{(n)}(r;\sigma) ,
\end{equation}
where
\begin{eqnarray}
I_{Q,q}^{(1)}(r;\sigma) & = & \int_{r_1<\sigma} {\rm d}^2 r_1
\sum_{q_1=\pm 1} \langle {\rm e}^{{\rm i}Qb\phi({\bf 0})}
{\rm e}^{{\rm i}q_1b\phi({\bf r}_1)} {\rm e}^{{\rm i}qb\phi({\bf r})} 
\rangle , \label{4.14} \\
I_{Q,q}^{(2)}(r;\sigma) & = & \int_{r_1<\sigma} {\rm d}^2 r_1
\int_{r_2<\sigma} {\rm d}^2 r_2 \sum_{q_1=\pm 1} \sum_{q_2=\pm 1} 
\nonumber \\ & &
\langle {\rm e}^{{\rm i}Qb\phi({\bf 0})}
{\rm e}^{{\rm i}q_1b\phi({\bf r}_1)} {\rm e}^{{\rm i}q_2b\phi({\bf r}_2)} 
{\rm e}^{{\rm i}qb\phi({\bf r})} \rangle , \label{4.15}
\end{eqnarray}
and so on.
As before, increasing the value of the guest charge $Q$ from $0$ to $\infty$,
the expansion (\ref{4.13}) represents the basis for the systematic
generation of ${\hat\sigma}$-corrections which are relevant (i.e., have
a negative exponent) in the ${\hat\sigma}\to 0$ limit.

In the stability region of the pointlike guest charge
$0\le Q < Q_0^*$, one has
\begin{equation} \label{4.16}
\langle {\rm e}^{{\rm i}Qb\phi({\bf 0})}
{\rm e}^{{\rm i}qb\phi({\bf r})} \rangle_{\sigma} \sim
\langle {\rm e}^{{\rm i}Qb\phi({\bf 0})}
{\rm e}^{{\rm i}qb\phi({\bf r})} \rangle
\quad \mbox{as ${\hat\sigma}\to 0$.}
\end{equation}
Inserting this relation together with the previously derived one (\ref{3.19}) 
into Eq. (\ref{4.12}) and repeating the procedure of Section 4.1, 
one ends up with the renormalized-charge formula (\ref{4.9}).

When $Q_0^*-\epsilon \le Q < Q_1^*$, the relevant contribution to 
the integral $I_{Q,q}^{(1)}(r;\sigma)$ (\ref{4.14}) comes from 
the $q_1=-1$ counterion correlation function.
Since the counterion position vector ${\bf r}_1$ is close to the origin
${\bf 0}$, we can apply the short-distance OPE formula
\begin{equation} \label{4.17}
{\rm e}^{{\rm i}Qb\phi({\bf 0})} {\rm e}^{-{\rm i}b\phi({\bf r}_1)}
\mathop{\sim}_{r_1\to 0} r_1^{-4 Q b^2} 
{\rm e}^{{\rm i}(Q-1)b\phi({\bf 0})} , 
\end{equation}
which implies
\begin{equation} \label{4.18}
I_{Q,q}^{(1)}(r;\sigma) \sim 2 \pi 
\frac{\langle {\rm e}^{{\rm i}(Q-1)b\phi({\bf 0})} 
{\rm e}^{{\rm i}qb\phi({\bf r})} \rangle}{m_1^{2-4Qb^2}} 
\frac{{\hat\sigma}^{2-4Qb^2}}{2-4Qb^2} \quad \mbox{as ${\hat\sigma}\to 0$.}
\end{equation}
The consideration of this formula in the basic expansion (\ref{4.13})
leads to
\begin{equation} \label{4.19}
\langle {\rm e}^{{\rm i}Qb\phi({\bf 0})} {\rm e}^{{\rm i}qb\phi({\bf r})} 
\rangle_{\sigma} \mathop{\sim}_{{\hat\sigma}\to 0} 
\langle {\rm e}^{{\rm i}Qb\phi({\bf 0})} 
{\rm e}^{{\rm i}qb\phi({\bf r})} \rangle 
- \frac{\pi z}{2 b^2} \frac{\langle {\rm e}^{{\rm i}(Q-1)b\phi({\bf 0})} 
{\rm e}^{{\rm i}qb\phi({\bf r})} \rangle}{m_1^{4b^2(Q_0^*-Q)}}
\frac{{\hat\sigma}^{4b^2(Q_0^*-Q)}}{Q_0^*-Q} .
\end{equation}
Close to the collapse point, $Q=Q_0^*-\epsilon$, using the large-distance
form-factor representation of two-point correlation functions (\ref{2.27})
and the formula (\ref{4.10}), singular terms of order $\epsilon^{-1}$
disappear from the rhs of Eq. (\ref{4.19}) in the same way as it was
in the case of $\langle {\rm e}^{{\rm i}Qb\phi({\bf 0})} \rangle_{\sigma}$,
see Eqs. (\ref{3.22})--(\ref{3.24}).
Finally, inserting (\ref{4.19}) into Eq. (\ref{4.12}) and using the
form-factor representation (\ref{2.27}) together with the result (\ref{3.22})
for $\langle {\rm e}^{{\rm i}Qb\phi({\bf 0})} \rangle_{\sigma}$, one obtains
\begin{equation} \label{4.20}
n_q(r) - n_q \mathop{\sim}_{r\to\infty} - n_q [q] g_Q^{(1)}(\sigma)
\lambda \left( \frac{\pi}{2 m_1 r} \right)^{1/2} \exp(-m_1 r) .
\end{equation}
Here,
\begin{equation} \label{4.21}
g_Q^{(1)}(\sigma) = \frac{N_Q^{(1)}(\sigma)}{D_Q^{(1)}(\sigma)}
\end{equation}
with
\begin{eqnarray}
N_Q^{(1)}(\sigma) & \displaystyle{\mathop{\sim}_{\hat\sigma\to 0}} &
\langle {\rm e}^{{\rm i}Qb\phi} \rangle [Q] - \frac{\pi z}{2 b^2} 
\frac{\langle {\rm e}^{{\rm i}(Q-1)b\phi} \rangle}{m_1^{4b^2(Q_0^*-Q)}}
\frac{{\hat\sigma}^{4b^2(Q_0^*-Q)}}{Q_0^*-Q} [Q-1] , \label{4.22} \\
D_Q^{(1)}(\sigma) & \displaystyle{\mathop{\sim}_{\hat\sigma\to 0}} &
\langle {\rm e}^{{\rm i}Qb\phi} \rangle - \frac{\pi z}{2 b^2} 
\frac{\langle {\rm e}^{{\rm i}(Q-1)b\phi} \rangle}{m_1^{4b^2(Q_0^*-Q)}}
\frac{{\hat\sigma}^{4b^2(Q_0^*-Q)}}{Q_0^*-Q} . \label{4.23} 
\end{eqnarray}
The denominator $D_Q^{(1)}(\sigma)$ is equal to the one-point average
$\langle {\rm e}^{{\rm i}Qb\phi({\bf 0})} \rangle_{\sigma}$ given by
Eq. (\ref{3.22}) and the numerator $N_Q^{(1)}(\sigma)$ possesses the
same structure as $D_Q^{(1)}(\sigma)$ but each term is multiplied by 
the function $[\cdots]$ (\ref{2.28}) with the argument related to
the phase of the corresponding averaged exponential field; we shall see
that this rule takes place also on higher levels.
At the collapse point $Q=Q_0^*$, the equality (\ref{4.10}) implies
that $g_{Q_0^*}^{(1)}(\sigma)\mathop{\sim}_{\hat\sigma\to 0} [Q_0^*]$,
and one recovers the pointlike result (\ref{4.2}). 
This means that $Q_{\rm ren}$ is the continuous function of the bare
charge $Q$ at the collapse point $Q_0^*$.
For $Q_0^*<Q<Q_1^*$, the term of order ${\hat\sigma}^{4b^2(Q_0^*-Q)}$ 
is dominant in the limit ${\hat\sigma}\to 0$ in both $D_Q^{(1)}(\sigma)$ 
and $N_Q^{(1)}(\sigma)$, so that $g_Q^{(1)}=[Q-1]$. 
The repetition of the procedure of Section 4.1, starting from Eq. (\ref{4.20}),
then leads to
\begin{equation} \label{4.24}
\beta Q_{\rm ren} = [Q-1] \lambda \left( \frac{\kappa}{m_1} \right)^2 ,
\quad Q_0^*\le Q < Q_1^* .
\end{equation}
Comparing this relation with Eq. (\ref{4.9}) (valid for $Q<Q_0^*$)
we see that, in the considered range of $Q$ values, the bare charge
$Q$ of the guest particle is effectively reduced by 1 due to the
condensation of one electrolyte counterion.

In the region $Q_1^*\le Q<Q_2^*$, one recovers the formula of type
(\ref{4.20}) with the substitution $g_Q^{(1)}(\sigma)\to g_Q^{(2)}(\sigma)$,
where $g_Q^{(2)}(\sigma)$ is again the ratio of type (\ref{4.21}):
\begin{equation} \label{4.25}
g_Q^{(2)}(\sigma) = \frac{N_Q^{(2)}(\sigma)}{D_Q^{(2)}(\sigma)} .
\end{equation}
Here, the denominator $D_Q^{(2)}(\sigma)$ is equal to the one-point 
average $\langle {\rm e}^{{\rm i}Qb\phi({\bf 0})} \rangle_{\sigma}$ 
given by Eq. (\ref{3.32}) and, as before, the numerator
\begin{eqnarray}
N_Q^{(2)}(\sigma) & \displaystyle{\mathop{\sim}_{\hat\sigma\to 0}} &
\langle {\rm e}^{{\rm i}Qb\phi} \rangle [Q] - \frac{\pi z}{2 b^2} 
\frac{\langle {\rm e}^{{\rm i}(Q-1)b\phi} \rangle}{m_1^{4b^2(Q_0^*-Q)}}
\frac{{\hat\sigma}^{4b^2(Q_0^*-Q)}}{Q_0^*-Q} [Q-1] \nonumber \\
& & + f_Q^{(2)} \frac{(\pi z)^2}{4 b^2} 
\frac{\langle {\rm e}^{{\rm i}(Q-2)b\phi} \rangle}{m_1^{8b^2(Q_1^*-Q)}}
\frac{{\hat\sigma}^{8b^2(Q_1^*-Q)}}{Q_1^*-Q} [Q-2] \label{4.26}
\end{eqnarray}
arises from $D_Q^{(2)}(\sigma)$ by multiplying each term by the function 
$[\cdots]$ with the argument related to the phase of the corresponding 
averaged exponential field.
In the ${\hat\sigma}\to 0$ limit, once again the term of order
${\hat\sigma}^{4b^2(Q_0^*-Q)}$ is dominant in both $D_Q^{(2)}(\sigma)$ 
and $N_Q^{(2)}(\sigma)$ in the considered range $Q_1^*\le Q<Q_2^*$,
so that $g_Q^{(2)} = [Q-1]$.
The previously derived formula (\ref{4.24}) thus applies to the whole
region $Q_0^*\le Q<Q_2^*$,   
\begin{equation} \label{4.27}
\beta Q_{\rm ren} = [Q-1] \lambda \left( \frac{\kappa}{m_1} \right)^2 ,
\quad Q_0^*\le Q < Q_2^* .
\end{equation}

It can be shown that the next term in Eq. (\ref{4.26}), of order 
${\hat\sigma}^{8 b^2(Q_1^*-Q)}$, becomes dominant in the region 
$Q_2^*\le Q<Q_4^*$. 
Consequently, in the limit ${\hat\sigma}\to 0$,
\begin{equation} \label{4.28}
\beta Q_{\rm ren} = [Q-2] \lambda \left( \frac{\kappa}{m_1} \right)^2 ,
\quad Q_2^*\le Q < Q_4^* .
\end{equation}
This relation describes the simultaneous condensation of two electrolyte
counterions onto the guest charge $Q$.

With the aid of the general analysis outlined at the end of Section 3.2,
one can extend the treatment, 
performed above in the range $Q_0^*\le Q < Q_2^*$, 
to an arbitrary interval $Q_{2(n-1)}^*\le Q < Q_{2n}^*$ $(n=1,2\ldots)$.
For a given $n$, the dominant term (in the ${\hat\sigma}\to 0$ limit)
in $\langle {\rm e}^{{\rm i}Qb\phi({\bf 0})} \rangle_{\sigma}$
is of type (\ref{3.40}).
The corresponding factor generated by 
$\langle {\rm e}^{{\rm i}(Q-n)b\phi} \rangle$ is $[Q-n]$, which implies
\begin{equation} \label{4.29}
\beta Q_{\rm ren} = [Q-n] \lambda \left( \frac{\kappa}{m_1} \right)^2 ,
\quad Q_{2(n-1)}^*\le Q < Q_{2n}^* 
\end{equation}
for $n=1,2,\ldots$. 
The bare charge $Q$ of the guest particle is therefore effectively
reduced by $n$ due to the condensation of $n$ electrolyte counterions.
Since $Q_{2n}^* = Q_c+n$, the formula (\ref{4.29}) tells us that
beyond the collapse point $Q_c$, i.e., when $\beta Q\ge 2$, 
the renormalized charge $Q_{\rm ren}$ is the periodic function of 
the bare charge $Q$ with period 1.
The minimum value of $Q_{\rm ren}$, given by
\begin{equation} \label{4.30}
\beta Q_{\rm ren}^{(\rm min)} = \lambda 
\sin\left( \frac{2\pi}{4-\beta} \right)
\frac{\pi\beta/(4-\beta)}{\sin^2[\pi\beta/(4-\beta)]} ,
\end{equation}
is acquired at the infinite sequence of points $Q_0^*,Q_2^*,Q_4^*,\ldots$;
the maximum value of $Q_{\rm ren}$, given by
\begin{equation} \label{4.31}
\beta Q_{\rm ren}^{(\rm max)} = \lambda 
\frac{\pi\beta/(4-\beta)}{\sin^2[\pi\beta/(4-\beta)]} ,
\end{equation}
is acquired at the infinite sequence of points $Q_1^*,Q_3^*,Q_5^*,\ldots$.
It is obvious that, in the considered limit ${\hat\sigma}\to 0$,
the renormalized charge does not saturate at a specific finite value
as $Q\to\infty$, but oscillates between the two extremes given by
Eqs. (\ref{4.30}) and (\ref{4.31}).
Note that $Q_{\rm ren}^{(\rm min)}$ is positive, i.e., has the sign
of the bare $Q$, in the considered Coulomb-gas stability region 
$0\le \beta<2$; consequently, there is no charge inversion in the system.

Like for example, the plot of $Q_{\rm ren}$ versus $Q$ at the inverse 
temperature $\beta=1$ is presented in Fig. 1. 
The solid line represents the true dependence given by Eqs. (\ref{4.9})
and (\ref{4.29}). 
The dashed line corresponds to the ``naive'' analytic continuation of 
the pointlike formula (\ref{4.9}) into the counterion-condensation 
region $Q\ge 2$.

\begin{figure}[h]
\begin{center}
\includegraphics{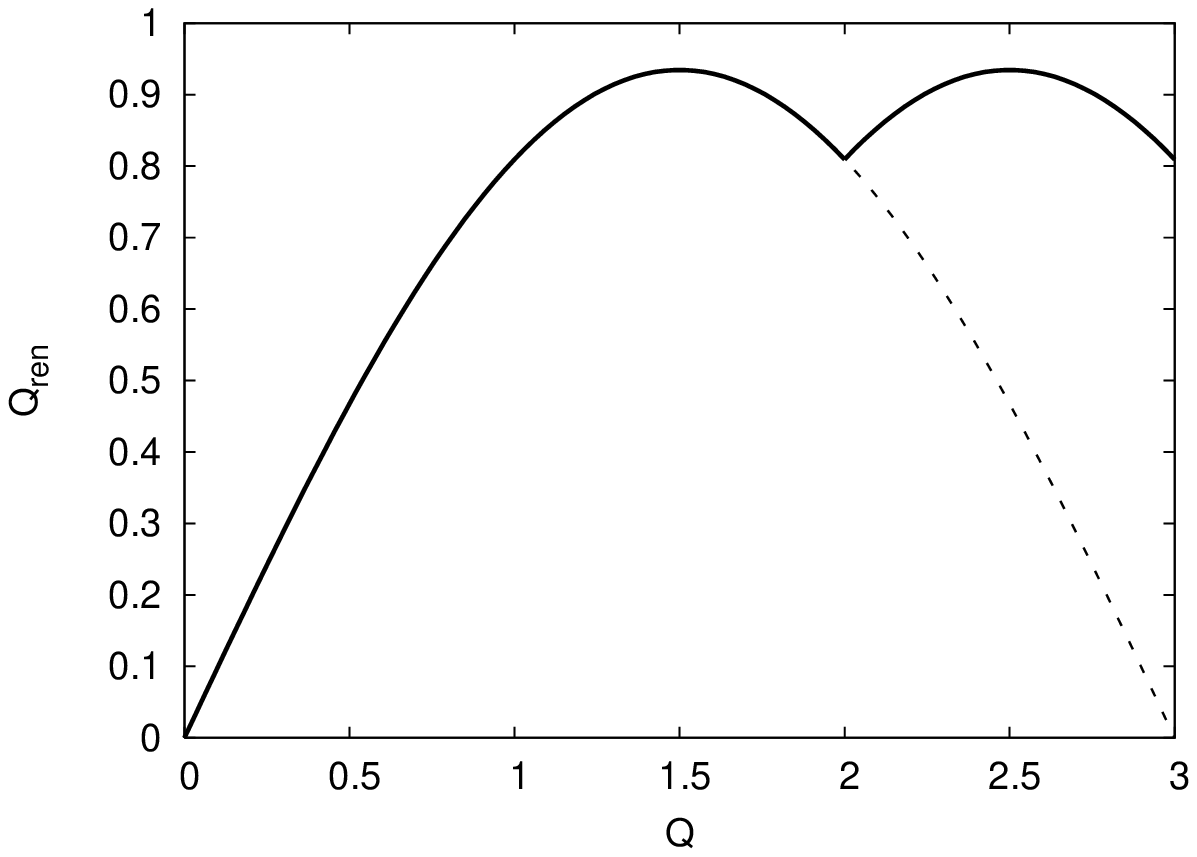}
\end{center}
\begin{caption}
{$Q_{\rm ren}$ versus the bare charge $Q$ at $\beta=1$ (solid line).}
\end{caption}
\end{figure}

We now check the obtained exact results on the high-temperature
PB scaling regime of limits $\beta\to 0$ and $Q\to\infty$ with
the product $\beta Q$ being finite.
In the guest-charge stability region $0\le \beta Q <2$, the consideration of 
$m_1\sim \kappa$, $\lambda\sim \beta$ and 
$[Q]\sim 4[\sin(\pi\beta Q/4)]/(\pi\beta)$ in Eq. (\ref{4.9}) results in
\begin{equation} \label{4.32}
\beta Q_{\rm ren} = \frac{4}{\pi}
\sin\left( \frac{\pi\beta Q}{4} \right) ,
\quad 0\le \beta Q < 2 .
\end{equation}
Evidently, $\beta Q_{\rm ren}$ is an increasing function of $\beta Q$
in the stability interval. 
It acquires its maximum value, equal to $4/\pi$, when $\beta Q$ approaches 
the collapse value $2$.
For $\beta Q \ge 2$, the two extreme values (\ref{4.30}) and (\ref{4.31})
coincide in the PB scaling regime:
\begin{equation} \label{4.33}
\beta Q_{\rm ren}^{(\rm min)} = \beta Q_{\rm ren}^{(\rm max)}
= \frac{4}{\pi} , \quad \beta Q \ge 2 .
\end{equation} 
This is equivalent to saying that $Q_{\rm ren}$ saturates at the value
given by $\beta Q_{\rm ren}^{(\rm sat)} = 4/\pi$ for all $\beta Q\ge 2$,
which is in agreement with the result (\ref{1.1}) of the Manning-Oosawa
theory of counterion condensation \cite{Tellez05a}.
We would like to emphasize that both the monotonic increase of
$\beta Q_{\rm ren}$ in the stability region $0\le \beta Q<2$ and
the saturation of $\beta Q_{\rm ren}$ at the uniform value for 
$\beta Q\ge 2$ are the specific features of the $\beta\to 0$ limit.
In the case of a strictly finite (nonzero) value of the inverse
temperature $\beta$, the previously described non-monotonic oscillatory 
dependence of $Q_{\rm ren}$ on the bare charge $Q$ takes place.

\renewcommand{\theequation}{5.\arabic{equation}}
\setcounter{equation}{0}

\section{Conclusion}
Exact results of ref. \cite{Samaj05b}, which concern thermodynamic
characteristics of a pointlike guest charge $Q$ (say, $\ge 0$) 
immersed in the infinite 2D Coulomb gas of $\pm 1$ charged species, 
are rigorously valid in the stability region of the guest particle 
$0\le \beta Q <2$.
The analytic continuation of these results to the ``collapse'' region
$\beta Q\ge 2$ turns out to be inadequate.
In the region $\beta Q\ge 2$, in order to prevent the direct collapse 
of electrolyte counterions onto the guest charge, the hard core of
radius $\sigma$, appearing in the dimensionless combination
${\hat\sigma}$ (\ref{3.12}), has to be attached to the guest charge
and only afterwards on can consider the limit ${\hat\sigma}\to 0$.
Each quantity related to the guest particle, evaluated in the presence 
of the hard core, exhibits some relevant ${\hat\sigma}$-corrections
with respect to its pointlike value which do not vanish, but on
the contrary diverge, in the limit ${\hat\sigma}\to 0$.
As $Q$ increases, the hierarchy of relevant ${\hat\sigma}$-corrections
becomes more complex and among all ${\hat\sigma}$-corrections
one is dominant in the sense that it fully determines the given quantity
in the limit ${\hat\sigma}\to 0$.
The main technicality of this paper consists in the systematic generation
of the relevant ${\hat\sigma}$-corrections as the value of $Q$ increases
and in the determination of the dominant ${\hat\sigma}$-correction.

The (excess) chemical potential of the guest charge, $\mu_Q^{\rm ex}$,
was the thermodynamic quantity of interest in Section 3.
In the case of the pointlike guest charge, $\mu_Q^{\rm ex}$ is expressible
in terms of the expectation value of the exponential field in the 2D
sine-Gordon model, see Eq. (\ref{2.12}).
The explicit result for this expectation value is available in the
whole guest-charge stability region $0\le \beta Q <2$, see
Eqs. (\ref{2.13}) and (\ref{2.14}).
The nontrivial extension of this result to the region $\beta Q\ge 2$
is possible due to the existence of the reflection relations
(\ref{3.4}) and (\ref{3.5}).
These reflection relations exhibit an infinite sequence of singularities
at the equidistant $Q$-points $\{ Q_n^* \}_{n=0}^{\infty}$ given
by Eq. (\ref{3.7}).
The chemical potential in the presence of the hard core around the
guest particle is the subject of Section 3.2, where the relevant 
${\hat\sigma}$-corrections to the pointlike $\mu_Q^{\rm ex}$
are systematically generated.
These ${\hat\sigma}$-corrections remove the artificial singularities
of the pointlike $\mu_Q^{\rm ex}$ at the points $\{ Q_n^* \}_{n=0}^{\infty}$,
which is an important consistency check of the generation procedure.
The hierarchy of the relevant ${\hat\sigma}$-corrections becomes more
and more complicated as the bare charge $Q$ increases, however,
the ${\hat\sigma}$-correction, dominant in the limit ${\hat\sigma}\to 0$,
is easily detectable for arbitrary-valued $Q$.

The crucial Section 4 is devoted to the evaluation of the renormalized
charge $Q_{\rm ren}$ of the guest particle.
The case of the pointlike guest charge is briefly reviewed in Section 4.1.
The inclusion of the hard core around the guest particle, studied in
Section 4.2, requires the application of both short-distance (the OPE
method) and large-distance (the form-factor method) asymptotic
expansions of two-point correlation functions in the 2D sine-Gordon model.
The systematic generation of the relevant ${\hat\sigma}$-corrections 
with respect to the pointlike renormalized charge is similar to the one 
outlined in Section 3.2 for the guest-charge chemical potential.
The final result for the renormalized charge (\ref{4.29}) is very simple.
In the problematic region $\beta Q\ge 2$, due to the condensation
of electrolyte counterions onto the guest charge, the renormalized
charge $Q_{\rm ren}$ is the periodic function of the bare charge $Q$
with period 1 and its value oscillates between the two extremes given
by Eqs. (\ref{4.30}) and (\ref{4.31}).
In the Poisson-Boltzmann scaling regime, these two extreme values
coincide and one recovers the standard Manning-Oosawa type of counterion
condensation with the uniform saturation value of $\beta Q_{\rm ren} = 4/\pi$ 
in the whole region $\beta Q\ge 2$.
From this point of view, the nature of the results in the PB scaling regime 
is fundamentally different from that of the exact result obtained at 
a strictly finite (nonzero) value of the inverse temperature $\beta$. 

\renewcommand{\theequation}{A.\arabic{equation}}
\setcounter{equation}{0}

\section*{Appendix}
The integral $J_Q^{(2)}$, defined in Eq. (\ref{3.30}), 
can be straightforwardly transformed to the form
\begin{eqnarray}
J_Q^{(2)} & = & 4 \pi \int_0^1 {\rm d}r\, r^{1+4b^2-4Qb^2}
\int_0^r {\rm d}r'\, (r')^{1-4Qb^2} 
\nonumber \\ & & \times \int_0^{2\pi} {\rm d}\varphi 
\left[ 1 - 2\cos\varphi \left( \frac{r'}{r} \right)
+ \left( \frac{r'}{r} \right)^2 \right]^{2b^2} . \label{A.1}
\end{eqnarray}
The expression in the square bracket can be expanded in the ratio
$(r'/r)$ by using Gegenbauer polynomials $\{ C_j^{\lambda}(t) \}$,
defined as the coefficients in the power-series expansion of the
function \cite{Gradshteyn}
\begin{equation} \label{A.2} 
\left( 1 - 2 t x + x^2 \right)^{-\lambda} = \sum_{j=0}^{\infty}
C_j^{\lambda}(t) x^j .
\end{equation}
Integrating then over $r'$ and $r$, one ends up with
\begin{equation} \label{A.3}
J_Q^{(2)} = \frac{\pi}{1+b^2-2Qb^2} \sum_{j=0}^{\infty} \frac{1}{2-4Qb^2+j} 
\int_0^{2\pi} {\rm d}\varphi\, C_j^{-2b^2}(\cos\varphi).
\end{equation}
Gegenbauer polynomials with the cosine argument are expressible as
\cite{Gradshteyn}
\begin{equation} \label{A.4}
C_j^{\lambda}(\cos\varphi) = \sum_{k,l=0\atop (k+l=j)}^j
\frac{\Gamma(k+\lambda) \Gamma(l+\lambda)}{k! l! [\Gamma(\lambda)]^2}
\cos(k-l)\varphi ,
\end{equation}
which implies
\begin{eqnarray}
\int_0^{2\pi} {\rm d}\varphi\, C_{2j+1}^{\lambda}(\cos\varphi)
& = & 0 , \label{A.5} \\
\int_0^{2\pi} {\rm d}\varphi\, C_{2j}^{\lambda}(\cos\varphi)
& = & 2 \pi \left[ \frac{\Gamma(j+\lambda)}{j! \Gamma(\lambda)}
\right]^2 . \label{A.6}
\end{eqnarray}
The consideration of these relations in Eq. (\ref{A.3}) leads to
\begin{equation} \label{A.7}
J_Q^{(2)} = \frac{\pi^2}{1+b^2-2Qb^2} \sum_{j=0}^{\infty} \frac{1}{1-2Qb^2+j} 
\left[ \frac{\Gamma(j-2b^2)}{j!\Gamma(-2b^2)} \right]^2 .
\end{equation}

Rewriting in Eq. (\ref{A.3}) the $j$th ratio $1/(2-4Qb^2+j)$ as the integral 
$\int_0^1 {\rm d}r\, r^{1-4Qb^2+j}$, the summation over $j$ 
in Eq. (\ref{A.7}) can be expressed as
\begin{equation} \label{A.8}
\sum_{j=0}^{\infty} \frac{1}{1-2Qb^2+j} 
\left[ \frac{\Gamma(j-2b^2)}{j!\Gamma(-2b^2)} \right]^2
= \frac{1}{\pi} \int_{r<1} {\rm d}^2 r\, r^{-4Qb^2}
\vert {\bf 1}-{\bf r} \vert^{4b^2} 
\end{equation}
The integral on the rhs of Eq. (\ref{A.8}) resembles the
Dotsenko-Fateev one $j_1(-Qb^2,b^2,b^2)$ [see the definition (\ref{2.20})], 
however, the domain of integration is restricted to the unit disc.
It is trivial to show that
\begin{equation} \label{A.9}
j_1(-Qb^2,b^2,b^2) = \int_{r<1}{\rm d}^2 r
\left[ r^{-4Qb^2} + r^{4(Qb^2-b^2-1)} \right]
\vert {\bf 1}-{\bf r} \vert^{4b^2} .
\end{equation} 
In the special case of equal exponents $-4Qb^2=4(Qb^2-b^2-1)$, i.e.,
when $Q= Q_c+1/2 \equiv Q_1^*$, one finds that
\begin{equation} \label{A.10}
j_1(-Q_1^*b^2,b^2,b^2) = 2 \int_{r<1} {\rm d}^2r\, r^{-4Q_1^*b^2}
\vert {\bf 1}-{\bf r} \vert^{4b^2} .
\end{equation} 
The comparison of this relation with Eq. (\ref{A.8}) implies
\begin{equation} \label{A.11}
\sum_{j=0}^{\infty} \frac{1}{1-2Q_1^*b^2+j} 
\left[ \frac{\Gamma(j-2b^2)}{j!\Gamma(-2b^2)} \right]^2
= \frac{1}{2\pi} j_1(-Q_1^*b^2,b^2,b^2) .
\end{equation}

\section*{Acknowledgments}
The support by a grant VEGA is acknowledged.









\end{document}